\RequirePackage{fix-cm}
\documentclass[twocolumn]{svjour3}          % twocolumn
\smartqed  % flush right qed marks, e.g. at end of proof
\usepackage{graphicx}
\usepackage{algorithmic}
\usepackage{graphicx}
\usepackage[sort&compress]{natbib}
\usepackage{amsmath,amssymb,amsfonts}
\usepackage{textcomp}
\usepackage{acronym}
\usepackage{color}
\usepackage{verbatim}
\usepackage{url}
\usepackage{xcolor}
\usepackage{multirow}
\usepackage{color, colortbl}
\usepackage{textcomp}

\begin{document}
\acrodef{LLM}{Large Language Model}
\acrodef{LLMs}{Large Language Models}
\acrodef{eLLMs}{Educational Large Language Models}
\acrodef{SLR}{Systematic Literature Review}
\title{Securing Educational LLMs: A Generalised Taxonomy of Attacks on LLMs and DREAD Risk Assessment %\thanks{Grants or other notes
}
% \subtitle{Do you have a subtitle?\\ If so, write it here}


\author{Farzana Zahid        \and Anjalika Sewwandi\and Lee Brandon \and  Vimal Kumar \and Roopak Sinha
        %etc.
}


\institute{Farzana Zahid \at
              University of Waikato, Hamilton, New Zealand. \\
                       \email{farzana.zahid@waikato.ac.nz}           %  \\
           \and
           Anjalika Sewwandi \at
              University of Waikato, Hamilton, New Zealand.\\
                       \email{ap623@students.waikato.ac.nz}   
              \and
           Lee Brandon \at
              University of Waikato, Hamilton, New Zealand. \\
                       \email{xl443@students.waikato.ac.nz} 
\and        Vimal Kumar \at
              University of Waikato, Hamilton, New Zealand. \\
                       \email{vimal.kumar@waikato.ac.nz}
    \and  
                       Roopak Sinha  \at
              Deakin University, Melbourne, Australia. \\
                       \email{roopak.sinha@deakin.edu.au}
}

\date{Received: date / Accepted: date}

\maketitle

\begin{abstract}
Due to perceptions of efficiency and significant productivity gains, various organisations, including in education, are adopting \ac{LLMs} into their workflows. Educator-facing, learner-facing, and institution-facing \ac{LLMs}, collectively, \ac{eLLMs}, complement and enhance the effectiveness of teaching, learning, and academic operations. However, their integration into an educational setting raises significant cybersecurity concerns. A  comprehensive landscape of contemporary attacks on \ac{LLMs} and their impact on the educational environment is missing. This study presents a generalised taxonomy of fifty attacks on \ac{LLMs}, which are categorized as attacks targeting either models or their infrastructure. The severity of these attacks is evaluated in the educational sector using the DREAD risk assessment framework. Our risk assessment indicates that token smuggling, adversarial prompts, direct injection, and multi-step jailbreak are critical attacks on \ac{eLLMs}. The proposed taxonomy, its application in the educational environment, and our risk assessment will help academic and industrial practitioners to build resilient solutions that protect learners and institutions.
\end{abstract}

\section{Introduction}
\label{sec:intro}
 \ac{LLMs} are designed for understanding and generating natural language text and solving complex tasks \cite{zhao2023survey,raiaan2024review,liu2023prompt,hadi2023survey,shen2023large,naveed2023comprehensive,kukreja2024literature}. These models utilise deep learning algorithms characterized by a vast number of parameters, and are trained on massive datasets to understand the relationship and trends among the linguistic constructs. With the advent of advanced \ac{LLMs} like PaLM \cite{chowdhery2022scaling}, LLaMA \cite{touvron2023llama}, Gemini \cite{team2023gemini}, Falcon \cite{Falcon}, GPT and its versions specifically GPT-3 \cite{GPT3} and GPT-4 \cite{GPT4}, DeepSeek \cite{deepseek}  and others \cite{kukreja2024literature},  these models 
mark a paradigm shift within numerous sectors. From finance to healthcare, and manufacturing to education, \ac{LLMs} are playing a major role in innovations, streamlining processes and redefining standards, and achieving human-level performance in applications like dialogue management, text
translation, and virtual assistance
\cite{bower2024should,hadi2023survey,raiaan2024review,shen2023large,liu2023prompt,pahune2023several}.

\acf{eLLMs} support learning in many
ways, such as personalised learning experiences across space and time, content generation, automatic grading, feedback, research assistance, scheduling, assessment evaluation, real-time problem solving, and other institutional support \cite{bower2024should,su151612451,mohammed2024chatgpt,xu2024large}. 
Glo-bal Market Insights has predicted that the AI education market, particularly using \ac{LLMs}, will reach \$20 billion by 2027 \cite{motlagh2024large,bahrini2023chatgpt}. Unfortunately, \ac{LLMs} have also sparked widespread cybersecurity concerns
in education \cite{mohammed2024chatgpt,atlam2025llms}. The increasing use of \ac{eLLMs} expands the attack surfaces and the entry points an attacker can use to compromise educational institutions \cite{motlagh2024large,su151612451}. Although education is not the sector most driven by financial gain, the large amount of personal data (student, employee, and institutional), intellectual property, research data, and lack of adequate security measures make them a target for cyber-attacks  \cite{sadiku2023cybersecurity,siphambili2024exploring}.  This is evident by the surge in the number of cyber-attacks on education recently \cite{incidents,incidents2024}. 

Ensuring security in educational workflows driven by \ac{eLLMs} is a continuous endeavour that requires dealing with the increasingly sophisticated cyber-attacks on \ac{eLLMs} models and their infrastructure. These attacks directly or indirectly impact the integrity of learning materials, public trust of educational institutions, privacy and security of staff, students, or associated stakeholders' information, academic operational continuity, and financial sustainability of educational institutions. Thus, it is crucial to understand the attack surfaces,  techniques, tactics, and potential attack vectors utilised by the attackers to ensure that \ac{eLLMs} operate as intended.

A scan of the current literature shows that despite the growing number of studies related to security issues in \ac{LLMs}, there is a need for in-depth analysis of attacks in terms of their level of sophistication, which could be helpful in understanding the evolving attack landscape and ascertaining effective defensive mechanisms. Furthermore, only a few of the existing works focus on risk analysis, while none emphasise the importance of evaluating risks related to critical sectors like education. 

% Risk analysis is an integral component in ensuring the security concerns of any organisation or sector. Different sectors have different security problems; risk analysis explores root processes and systems deeper to uncover those problems. Therefore, identifying the security risks in education sector that rely on \ac{LLMs} can help with targeted investment to lift productivity. %enables them to allocate the resources where they are considerably required, enhancing their overall productivity and security.     

 \textit{This article introduces a generalised taxonomy of cyber-attacks on \ac{LLMs} and analyses the criticality (seve-rity) of the identified attacks in education}. We conduct a \ac{SLR} \cite{KITCHENHAM20097} to identify the current security issues within \ac{LLMs}. This \ac{SLR} study explores the following research questions. 

\begin{itemize}
\item RQ1 What are the key security attacks on \ac{LLMs}?
\subitem -RQ1.1 How can the attacks identified in RQ1 be characterised by the level of sophistication? 
\subitem -RQ1.2 How can the characterisation, resulting from RQ1.1, assist in recognising critical attack vectors and general impact of attacks on  \ac{LLMs}?  
  
\item RQ2  How could the attacks identified in RQ1 be evaluated and prioritised within the education sector? 
\end{itemize} We then propose key classification criteria to
categorise cyber-attacks on \ac{LLMs} based on their level of complexity/sophistication, an area that has not yet been explored by any existing studies. We also analyse the attack vectors and their impacts. This proposed taxonomy categorising the sophisticated attacks on \ac{LLMs}, attack vectors, and impacts will be useful equally for academic and industrial practitioners to secure \ac{eLLMs}. As presented in Section~\ref{sec:application}, another novel contribution of this study is to quantify the risks associated with attacks on \ac{eLLMs} using the DREAD (Damage, Reproducibility, Exploitability, Affected Users, and Discoverability) risk assessment framework \cite{threatmodel,naik2024comparative}. 
The DREAD framework provides an elegant, systematic, and flexible approach to identifying the critical security risks to an organization. Its criteria are independent (not correlated with each other), and straightforward in both application and interpretability, making it suitable for our study to identify and address high-priority \ac{eLLMs} security risks quickly before they can be exploited, resulting in optimal business and technical impact \cite{smith2016car}. Applying DREAD scores for aspects of the education sector where the adoption of \ac{eLLMs} could mean higher risk allow professionals to proactively fortify the corresponding security postures.  

% to perform the threat analysis ...  and identified the appropriate security controls. 
% To the best of our knowledge, no existing work . In this study, we have utilised the DREAD risk analysis model to ....  . Thus, the motivation of this work leads to the following research questions:  
\

\begin{table*}[]
\caption{Comparison of our Work with the Existing Secondary Studies by Research Type (Survey, Exploratory Study, Empirical Study, Opinion Paper), LLMs Types (General, ChatGPT, Gemini or others), Attack Taxonomy (Yes/No), Attack Vectors (Yes/No), Attack Impact (Yes/No), Risk Analysis (Method/No)}
\label{survey}
\begin{tabular}{|l|l|l|l|l|l|l|}
\hline
\textbf{References} &
  \textbf{\begin{tabular}[c]{@{}l@{}}Type of \\ Research\end{tabular}}  & \textbf{\begin{tabular}[c]{@{}l@{}}Types of \\LLMs\end{tabular}}  &
  \textbf{\begin{tabular}[c]{@{}l@{}}Attack \\ Taxonomy\end{tabular}} &
  \textbf{\begin{tabular}[c]{@{}l@{}}Attack \\ Vectors\end{tabular}} &
 \textbf{\begin{tabular}[c]{@{}l@{}}Attack \\ Impact \end{tabular}} &
   \textbf{\begin{tabular}[c]{@{}l@{}}Risk \\ Analysis\end{tabular}} \\ \hline
 \cite{YAO2024100211} & Survey  & General  & No & No & No  & No \\ \hline
\cite{shayegani2023survey} & Survey & General & No & No  & Yes & No \\ \hline
\cite{shen2023large} & Survey & General & No & Yes & Yes   &  No\\ \hline
 \cite{iqbal2023llm} & Survey & ChatGPT &Based on platform plugins &Yes & No    & No   \\ \hline
\cite{yang2024comprehensive}& Survey  & General & Based on backdoor Attacks  & Yes & No & No  \\ \hline
\cite{wu2023unveiling} & Survey & ChatGPT & No & No & Yes & No \\ \hline
\cite{pankajakshan2024mapping} & Survey& General & No &  No & Yes & OWASP \\ \hline 
\cite{hadi2023survey} & Survey & General & No& No & No  & No  \\ \hline 
\cite{derner2023beyond} & Empirical Study & ChatGPT & No & No & Yes  & No \\ \hline 
\cite{sebastian2023chatgpt} & Exploratory Study & ChatGPT & No & No & Yes & No \\ \hline
\cite{cui2024risk} & Survey & General & No & Yes & No  & \begin{tabular}[c]{@{}l@{}} Benchmark \\datasets \end{tabular}  \\ \hline
\cite{chu2024comprehensive} & Survey & General & Based on jailbreak & Yes &  No & No \\ \hline
\cite{chowdhury2024breaking} & Survey & General & \begin{tabular}[c]{@{}l@{}}Based on prompt injection, \\jailbreak,data poisoning \end{tabular}& No &  Yes & No \\ \hline 
\cite{zhang2025llms} & Survey & General & \begin{tabular}[c]{@{}l@{}}No (brief discussion on prompt\\injection, jailbreak,backdoor)\end{tabular}& No &  No & No \\ \hline 
\textit{Our work} & \textit{Survey} & \textit{General} & \textit{Based on Attack Complexity} & \textit{Yes} & \textit{Yes}  & \textit{DREAD} \\ \hline

\end{tabular}
\end{table*}

\textit{Thus, the primary contributions of this study include:}
\begin{enumerate}
\item A systematic literature review of the up-to-date security attacks on \ac{LLMs}, presented in Section~\ref{sec:1}.
    \item  A generalised attack taxonomy on \ac{LLMs},  based on their level of complexity, detailed in Section~\ref{sec:tax} and Section~\ref{sec:application}.
    \item Analysis of the attack vectors and their potential impact to identify the most critical threats to \ac{LLMs}-based workflows, presented in Section~\ref{sec:application}.
    \item Mapping the proposed \ac{LLMs}-based attack taxonomy to the education sector and quantification of the associated risks using the DREAD model, given in Section~\ref{sec:application}. 
\end{enumerate}

% The rest of this article is organised as follow. Sec.~\ref{sec:1} describe \ac{SLR}. Sec.~\ref{sec:tax} and Sec.~\ref{sec:application}  present the proposed taxonomy, and its application in eduction using DREAD model. Concluding remarks and future directions are presented in Sec.~\ref{sec:conclusion}.

\section{Systematic Literature Review}\label{sec:1}

A \acf{SLR} conducted includes the following phases: planning, conducting and reporting the review \cite{KITCHENHAM20097}.  The Covidence tool was used to ensure clear reporting for a systematic review \cite{kellermeyer2018covidence}. 

\subsection{Planning} \label{sec:plan}

\textbf{Scope definition and formulation of research que-stions:} To answer RQ1, a comprehensive literature review  is conducted to identify and scrutinise existing research works within the area of \ac{LLM} security. The scope of this research is to investigate and analyse security attacks on \ac{LLMs}. We propose an easy-to-understand generic taxonomy of attacks on \ac{LLMs} based on the level of attack sophistication (attack complexity), analyse the various attack vectors for each identified attack, and determine the impact of those attacks. Moreover, we also quantify the risks posed by each attack in the education sector. Our \ac{SLR} also identifies several secondary studies in this area \cite{YAO2024100211,yang2024comprehensive,shen2023large,hadi2023survey,shayegani2023survey,wu2023unveiling,pankajakshan2024mapping,derner2023beyond, chowdhury2024breaking,sebastian2023chatgpt,iqbal2023llm,cui2024risk}, but these studies differ significantly in their focus and methodology, as shown in Table~\ref{survey}. 

% We propose an easy to understand taxonomy of attacks on \ac{LLMs} based on the level of attack sophistication (attack complexity), analyse the various attack vectors for each identified attack and determine the impact of those attacks. Moreover, we also assess the risk poses by each attack in education sector and none of the existing studies integrate this perspective in their findings. 

Based on the scope of our study, we formulated the research questions mentioned in the Section ~\ref{sec:intro}.
\label{sec:literature}

\noindent\textbf{Database selection and search query:} IEEE Xplore, SpringerLink, and Scopus were selected for this study. Scopus, the largest commercially accessible database of peer-reviewed articles, also encompasses IEEE Xplore and SpringerLink. Nonetheless,
we conducted individual searches of all these databases to ensure completeness.

For filtering out the primary studies, we follow the methods reported in \cite{petersen2015guidelines}, and identified the following security-related keywords - \texttt{secur*}, \texttt{attack}, \texttt{threat}, \texttt{vuln*}, 
and \texttt{risk}. For \ac{LLMs}, we used terms like \texttt{large language models}, and \texttt{LLM}. The following query string in the Scopus format represents the final combination of the above keywords/phrases used in our SLR: 

\begin{verbatim}
(TITLE-ABS-KEY (large AND language AND 
model OR "LLM" OR "Large language model") 
AND TITLE-ABS-KEY (secur*) 
OR TITLE-ABS-KEY (threat) OR
TITLE-ABS-KEY (vuln*) OR 
TITLE-ABS-KEY (risk)) 
AND PUBYEAR > 2019 AND PUBYEAR < 2025
\end{verbatim}

\noindent\textbf{Inclusion/Exclusion criteria:} The inclusion criteria (\texttt{INC}) and exclusion criteria
(\texttt{EXC}) that were utilised for the selection of only the relevant studies from the search results are represented in Table~\ref{criteria}.
\begin{table}[!h]
\caption{ Inclusion and Exclusion Criteria}
\label{criteria}
\scriptsize
\begin{tabular}{|p{0.04\columnwidth}|l|p{0.7\columnwidth}|}
\hline
\multirow{6}{*}{\rotatebox[origin=c]{90}{\begin{tabular}[c]{@{}l@{}}     \textbf{ Inclusion} \\ \textbf{Criteria} \end{tabular}}} &
  INC1 &
  Studies that investigated \ac{LLMs} security issues. \\ \cline{2-3} 
 &
  INC2 &
  Studies that discuss the concepts like open challenges, problems related to the security issues within \ac{LLMs} \\ \cline{2-3} 
 &
  INC3 &
Studies published in conferences, journals, technical reports, pre-prints (as most of the recent articles are shared as pre-prints) \\ \cline{2-3} 
 & INC4  & Research studies that appeared since 2020 till now                                      \\ \cline{2-3} 
 & INC5  & Studies that focused on the real-world applications of \ac{LLMs}       \\ \cline{2-3} 
 & INC6  & Studies that include at least one of the specified keywords                             \\ \hline
\multirow{8}{*}{\rotatebox[origin=c]{90}{\begin{tabular}[c]{@{}l@{}}     \textbf{ Exclusion} \\ \textbf{Criteria} \end{tabular}}} &
  EXC1 &
 \begin{tabular}[c]{@{}l@{}}Studies where title, keywords and/or abstract \\do not lie within defined scope\end{tabular} \\ \cline{2-3} 
 & EXC2 & Studies do not investigating any security issues within \ac{LLMs}      \\ \cline{2-3} 
 & EXC3 & Studies that address solely the concept of privacy issues of \ac{LLMs} \\ \cline{2-3} 
 & EXC4 & Studies focusing on attacks that could be launched using \ac{LLMs}    \\ \cline{2-3} 
 & EXC5 & Studies that do not have full text                                                      \\ \cline{2-3} 
 & EXC6 & Books, thesis, tertiary studies, tutorial and opinion papers                            \\ \cline{2-3} 
 & EXC7 & Studies not written in English                                                          \\ \cline{2-3} 
 & EXC8 & Studies whose new version is available or are not peer-reviewed                         \\ \hline
\end{tabular}
\end{table}

\subsection{Conducting the review} \label{sec:conduct}
\textbf{Search and data extraction:} 
For this study, we performed both automated (using Covidence) and manual searching. 

\begin{figure}[!h]
    \centering
    \includegraphics[width=\columnwidth]{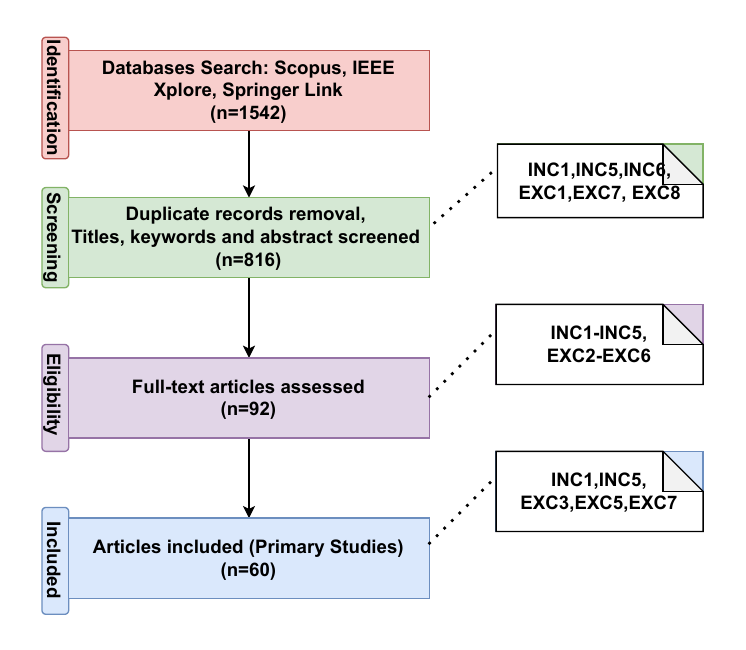}
    \caption{Search execution chronology}
     \label{fig:slr}
       \end{figure}

 Figure~\ref{fig:slr} shows the search execution chronology (Covidence's PRISMA flow diagram). The initial automated search resulted in 1542 articles. Due to duplicate records and screening of the titles, keywords, or abstracts, a significant number of studies were excluded, leaving 816 studies for eligibility selection. A further 724 studies were removed after carefully examining each study's introduction, conclusion, and full text. The number dropped to 60 after meta-analyses. Data extraction from automatic search includes the identification of the keywords by reading abstracts, introductions, and conclusions (if needed) \cite{petersen2015guidelines}.  
 In the case of manual search, ATLAS \cite{atlas}, AI Incident Database \cite{AIincident} and OWASP framework \cite{LLMsecurity}, and white papers related to the security issues of \ac{LLMs} were selected. As there is a continuous rise in the number of security-related studies on \ac{LLMs}, manual search plays an important role in enhancing the confidence of the comprehensiveness of the review. For manual search, data extraction is performed using the keywords selected for our query string. Furthermore, scanning the manual results for the attack scenarios and extracting the utilised tactics, techniques, and sub-techniques for particular attacks on \ac{LLMs} resulted in including ten more relevant articles.   
 
 % Thus, the insights and analysis of the existing studies (identified manually or automatically) help us to develop a taxonomy of the attacks on \ac{LLMs}, discussed in Sec.~\ref{sec:tax}.
\ 

\noindent\textbf{Quality Assessment:} Our quality assessment criteria contains five measures QC1–5 \cite{khan2019landscaping}. Each study is assigned a score of 0, 0.5, or 1 for each measure. The final quality score for a study is the aggregation of the individual scores, which is quantified as low when 0.5 $\leq$ quality score $\leq$ 2, medium for 2.5 $\leq$ quality score $\leq$ 3, or high (3.5 $\leq$ quality score $\leq$ 5). 

Based on the measures mentioned above, the scoring for our \ac{SLR} is calculated as follows.

\noindent\texttt{QC1.} Our study clearly states the objective of the research, and so it gets a score of 1.

\noindent\texttt{QC2.} A score of 1 is assigned as Inclusion and Exclusion criteria are defined.

\noindent\texttt{QC3.} A score of 1 for presenting an explicit synthesis method based on a well-used methodology.

\noindent\texttt{QC4.} The quality assessment of selected primary studies was performed, but
not reported, so our study gets a score of 0.5. 

\noindent\texttt{QC5.} A score of 1 for providing information about each primary studies.

Overall, our \ac{SLR} scores 4.5, indicating it is a high-quality review.

\subsection{Reporting (threats to validity)} \label{sec:valid}

\noindent\textbf{Overlooking Important Relevant Studies}: An \ac{SLR} is intended to cover the depth of a research area by analysing the existing works in that area. There is a chance of overlooking some relevant current literature, so the query string is formulated to retrieve the maximum number of studies from the databases. The titles and abstracts of the articles could also be ambiguous. Therefore, we thoroughly read the introductions, conclusions, and full-texts, if needed, to ensure the inclusion of relevant primary studies. Moreover, selecting the specific databases could also result in missing studies; hence, a manual search was also performed to mitigate this threat. We also used the Covidence tool for \ac{SLR}. The use of the tool and manual search therefore provides another layer of assurance and helps find articles that may be missed due to the use of non-standard terminology.

\noindent\textbf{Researcher Bias}: Researcher bias could impact the validity of research. The systematic literature review protocol was established and followed carefully with the support of domain experts and co-authors.  

\noindent\textbf{Selection of the Query String}: The final selection of the primary studies depends on the scope, novelty of research areas, and search strings. To create a query string that could not be very strict and define the scope, we tweaked it to remain comprehensive (return all relevant papers) while reducing the number of irrelevant papers returned. For example, with the keyword \texttt{risk} in the query string, we omitted keywords like risk \texttt{analy*}, \texttt{mitig*} and \texttt{assess*} intentionally because even by including these terms, the result of retrieving the number of studies does not change. Also, including keywords such as language model, natural language processing, natural languages, deep learning, machine learning, and Generative AI reduces the number of articles retrieved to only 152. Similarly, the keyword related to LLMs in education results in numerous irrelevant papers, while the other relevant papers were already filtered out using our final query mentioned in  Section~\ref{sec:plan}. 

\begin{figure*}[]
   \centering
   \caption{A Generalised Taxonomy of Attacks on LLMs}
\includegraphics[width=0.8\textwidth]{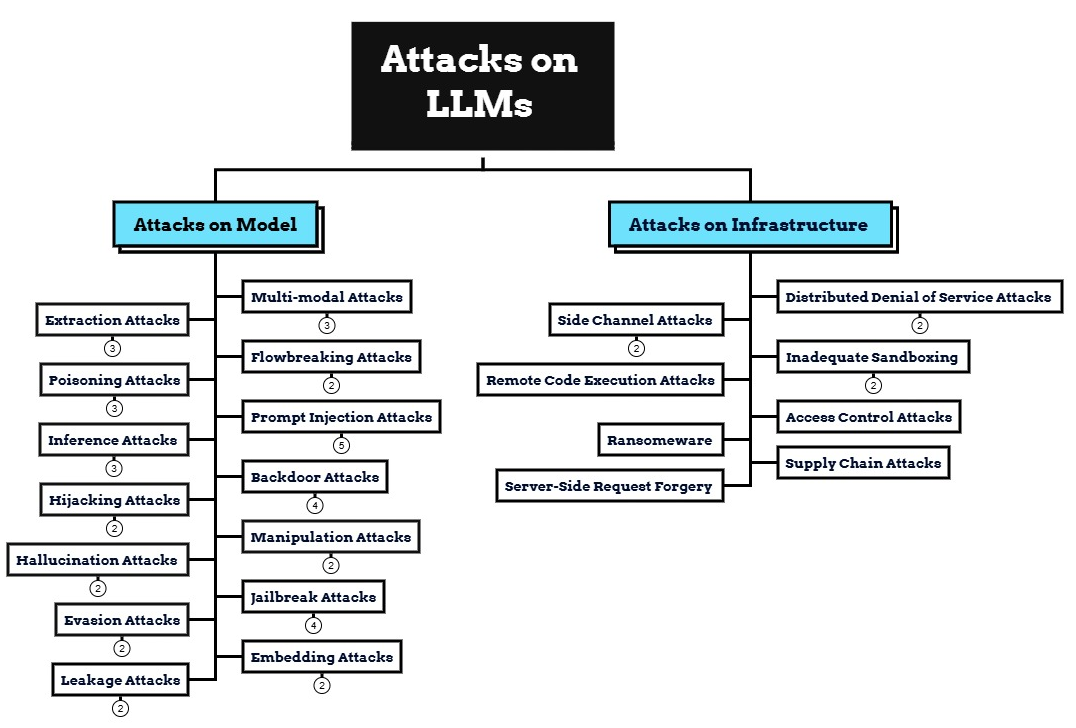}
  \label{fig:maintax} 
\vspace{-15pt}
     \end{figure*}
\section{Taxonomy of Security Attacks on Large Language Models (LLMs)}
\label{sec:tax}
% The following section discusses the results and findings of the \ac{SLR}. 
 \ac{LLMs} are susceptible to various attacks (Figure~\ref{fig:maintax}), which are either on their models (parameters, hyperparameters, model input, test data, training data, model documentation) directly or their infrastructure (deployments, storage, network, servers, hardware). We propose a novel taxonomy to classify the identified attacks on models and associated infrastructure based on the \textit{attack complexity} or sophistication level. 
 
  \ 
  
 \noindent\textbf{Sophistication Level (Attack Complexity}): Attack complexity or the level of the sophistication indicates the extent of the actions that must be taken by the attacker to compromise \ac{LLMs}. The actions are based on the need of specialised skills, knowledge of the model and/or infrastructure or ease of exploitation. We used the following attack complexity metric to categorise the attacks on \ac{LLMs} mentioned in the selected primary studies.

\begin{figure*}[!h]
   \centering
   \caption{A Breakdown of Taxonomy of Attacks on \ac{LLMs} Models}
\includegraphics[width=0.8\textwidth]{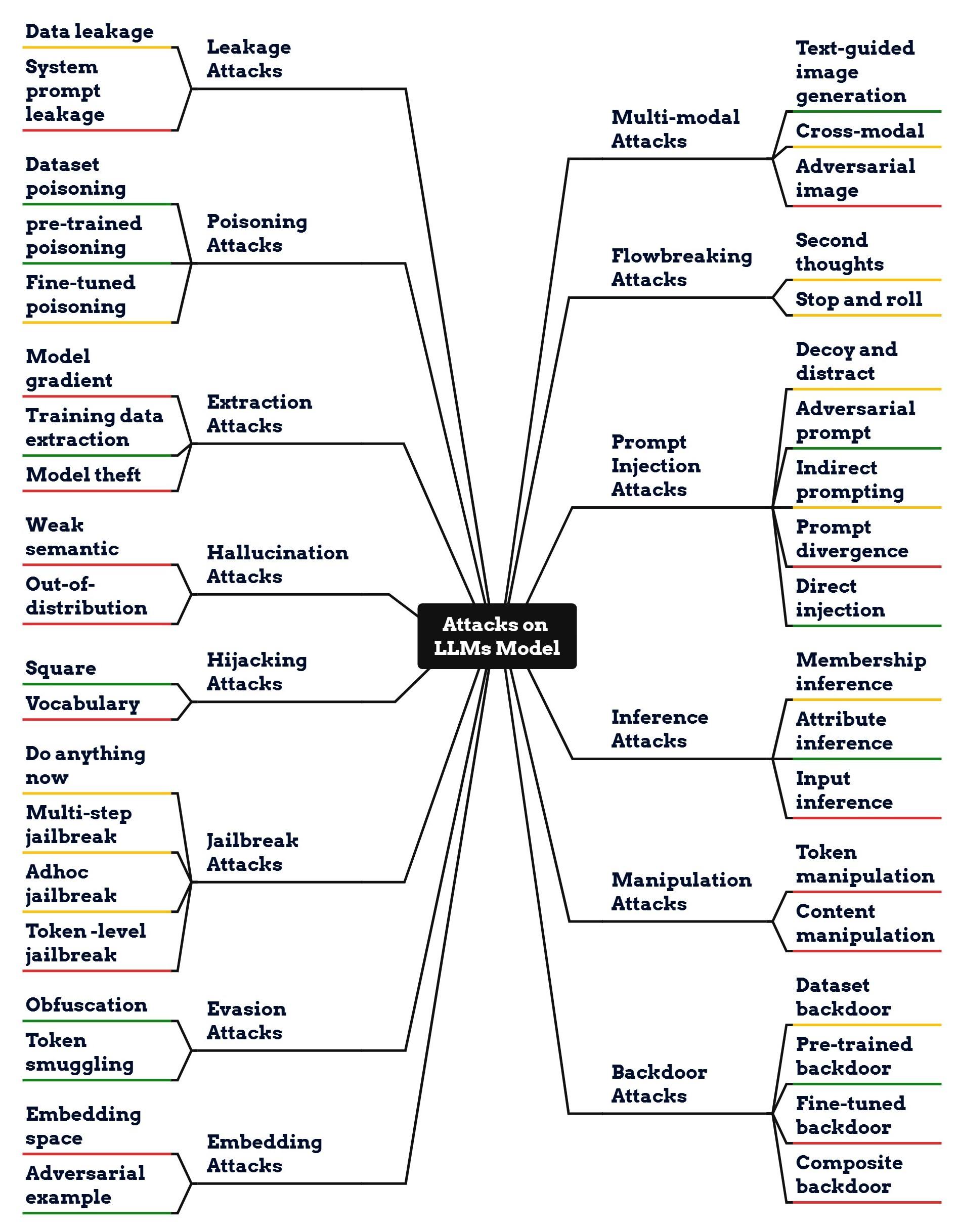}
  \label{fig:flow2} 
%\vspace{-15pt}
     \end{figure*}

\begin{figure*}[!h]
   \centering
   \caption{A Breakdown of Taxonomy of Attacks on LLMs Infrastructure}
\includegraphics[width=0.75\textwidth]{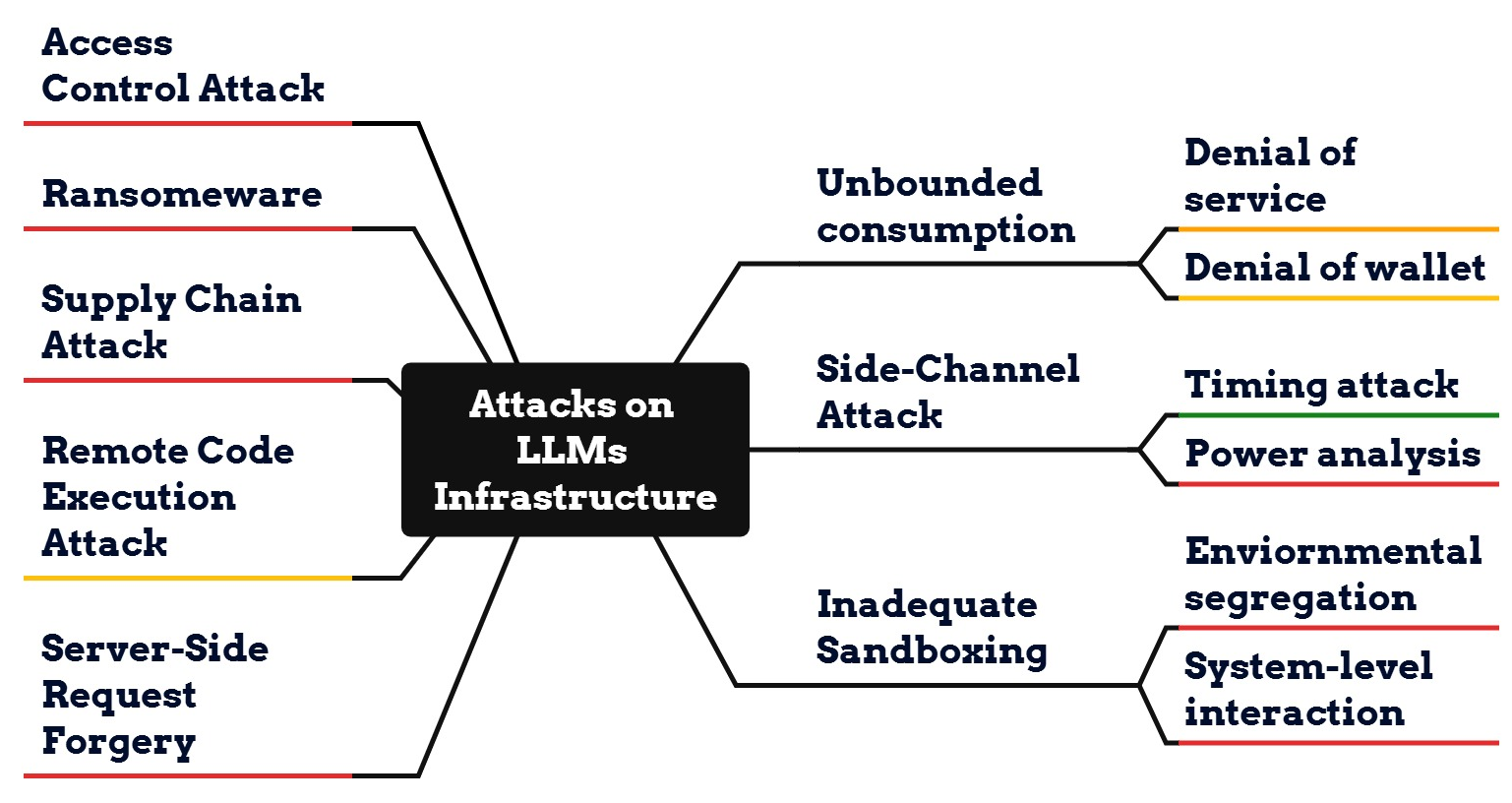}
  \label{fig:infr} 
%\vspace{-15pt}
     \end{figure*}

1. \textbf{High (H)}: The level of sophistication is \textit{high} if the attacker has to use specialised or advance tactics, skills,  methods or tools to compromise the \ac{LLMs} or needs an in-depth knowledge of the \ac{LLMs} model or infrastructure. These attacks are indicated by the color red in Figure~\ref{fig:flow2} and Figure~\ref{fig:infr}. 

2. \textbf{Medium (M)}: The attack sophistication level is \textit{medium} when the attacker performs multiple steps in a sequence without using very specialised techniques or tools to compromise the \ac{LLMs}. Medium level attacks are represented by yellow in Figure~\ref{fig:flow2} and Figure~\ref{fig:infr}. 

3. \textbf{Low (L)}: The level of sophistication is \textit{low} (illustrated in green color in Figure~\ref{fig:flow2} and Figure~\ref{fig:infr}) if the attacker uses single step (maliciously craft simple or direct prompt), or require less expertise or does not require the internal knowledge of the model or the infrastructure to launch an attack on \ac{LLMs}. 
% The model include parameters, hyperparameters, model input, test data, training data, model documentation and process of their development including experiments. 
% The detail of each attack, illustrated in Figure~\ref{fig:flow2} and Figure~\ref{fig:infr}, is in the following sections.

The detail of each type of attack  is presented in following sections. The summary of attack vectors and impacts is given in Table~\ref{vectors}.

% The models include parameters, hyperparameters, model input, test data, training data, model documentation and process of their development including experiments 
% .Due to different characteristics of the attacks, utilised attack vectors, the 
\subsection{Security Attacks on LLMs Models} \label{sec:taxmodel}

This section discusses various security attacks on the LLM models, as presented in Figure~\ref{fig:flow2}, along with their corresponding vectors and impacts. Table~\ref{Complexity} further elaborates on the Figure~\ref{fig:flow2} by presenting the complexities of each identified attack, determined using the previously mentioned sophistication level metric. Also, Table~\ref{vectors} summarises the associated vectors and the specific impacts of each attack. 

\ 

\noindent\textbf{Multi-Modal Attack}: A multi-modal attack on \ac{LLMs} is an adversarial attack to exploit the processing and understanding capabilities of \ac{LLMs} when dealing with different input types \cite{LLM2024,zhang2025jailguard}. To launch this attack, the attacker uses various attack vectors such as adversarial images, crafted noise or text with images, poisonous association to manipulate the \ac{LLMs}' generated output or their operational behaviours \cite{shayegani2023survey}. These attacks could take many forms like text guided image generation, cross-modal attack and adversarial image. 

In \textit{text guided image generation} attack, an intruder embeds malicious text with an image to trigger the \ac{LLMs} for the generation of the manipulated image with malicious intent like spreading misinformation \cite{kou2023character,wu2023bloomberggpt}. The attacker could either use pre-trained models or should have a knowledge of basic prompt engineering and with trial and error, he can crafts prompts to generate misleading output. Therefore, the attack complexity of this attack is low (L). 

In \textit{cross-modal attacks}, attackers exploit vulnerabilities in the interaction between different modalities, such as, text, audio, images, etc. to confuse the model \cite{wu2023bloomberggpt}. An attacker should have some knowledge of the model such as, pertaining to the linkage of text and images, and needs knowledge of prompt engineering to introduce minor inconsistencies in the modalities.  Also, attackers need multi-step procedures to launch this attack, first they exploit the vulnerability of one modality (such as audio)  and then utilise that compromised modality to trigger a specific response from the \ac{LLM}. Therefore, the attack complexity of this attack is medium (M). 

\textit{Adversarial image} attacks involve subtle modifications to images using hidden patterns such as imperceptible noise or minor pixel-level adjustments that mislead \ac{LLMs} to trigger the specific response \cite{zhang2024adversarial,shayegani2023survey,wu2024adversarial}. The attack complexity is high (H) because the attacker adds adversarial perturbations in the images, so they need high expertise and in-depth knowledge of the model to launch this attack.     

\ 

\noindent\textbf{Flowbreaking Attack}: Flowbreaking is a newly introduced/novel \ac{LLM} attack that targets the reasoning and coherence of \ac{LLM} models while generating the response \cite{suicide_bot_ai_attack}. Compared to input data manipulation attacks, the internal logic of the model's output is disrupted by flow-breaking attacks. Even benign prompts can lead the model to produce incorrect or harmful responses or result in information disclosure. There are two types of flowbreaking attack: second thoughts, and stop and roll \cite{generative_ai_under_attack,suicide_bot_ai_attack}.  

A \textit{second thoughts} attack occurs when \ac{LLMs} 
models initially provide a response to the prompt but halt or retract upon detecting a sensitive topic and either generate a simple error message or new modified content; attackers exploit this behavior to extract sensitive information. This attack requires prompt engineering skills and some knowledge of the model to exploit the guardrails (filters) or streaming window. Therefore, the attack complexity is medium (M).   

A \textit{stop and roll} attack involves the manipulation of \ac{LLM} output using hidden commands or crafting specific prompts. During the answering phase, the attacker breaks the flow of the \ac{LLM}'s reasoning by pressing the stop button, but the answer is still streaming and cannot be deleted. This attack results in unauthorized actions, information disclosure, or potentially damaging responses even though the system policies are violated. The ease of exploitation is simple using the stop button. However, the attacker needs model knowledge and some expertise to craft the specific instructions, so the complexity of the attack is medium (M). 

\ 

\noindent\textbf{Prompt Injection Attack}: In the prompt injection attack, malicious prompts replace the LLM's original instructions, manipulating them to respond to different queries rather than fulfill their intended function 
\cite{liu2023prompt,LLMsecurity2025,peng2024securing,alla2025cyberattacks}. There are various forms of prompt injection attacks, such as adversarial prompt injection attacks, decoy and distraction prompt injection attacks, indirect prompt injection attacks, prompt divergence attacks, and direct prompt injection attacks.

An \textit{adversarial prompt injection} attacker exploits the model's instruction-following behaviour to mislead LLM's intended response by directly adding adversarial instructions. Adversarial prompt injection, thus, can result in undesirable or unauthorised outputs, such as offensive responses or unauthorised data disclosure \cite{schulhoff2023ignore}. Attack vectors are user input fields or APIs that allow external text input to the \ac{LLMs}. In this attack, attackers need basic prompt engineering skills to craft effective phrases without requiring detailed knowledge of the model's internal workings. The adversarial prompt injection attacks are straightforward due to \ac{LLMs}' tendency to follow prompts precisely \cite{schulhoff2023ignore}. Thus,  the attack complexity is low (L). 

Likewise, in \textit{decoy and distraction prompt injection attacks}, the attacker misguides the \ac{LLMs} by embedding off-topic or confusing information within the prompt \cite{gupta2023chatgpt}. This decoy information causes \ac{LLMs} to prioritise the distraction, shifting the model's attention from the main question to the misleading content, leading to inaccurate or off-topic outcomes \cite{schulhoff2023ignore}. The attacker must be familiar with prompt structuring and model's prioritising to place distractions into the input. As a result, decoy and distraction prompt injection attacks have medium complexity (M) because they require an understanding of prompt dynamics to mislead the model accurately \cite{gupta2023chatgpt}. The attack vectors are frequently seen in chat interfaces or information retrieval systems where off-topic details can be embedded easily \cite{schulhoff2023ignore}.

In an \textit{indirect prompt injection attack}, adversarial prompts are inserted into retrievable data sources that the \ac{LLMs} access \cite{greshake2023not,zhan2025adaptive}. This causes the model to process compromised data, which may result in unintended responses or data leaks. Because these indirect prompt injection attackers exploit the model's data retrieval processes, \ac{LLMs} unknowingly execute embedded commands from external sources, allowing the attacker to remotely manipulate outputs without requiring direct prompt input. Indirect prompting leads to excessive agency vulnerability \cite{LLMsecurity2025}. Thus, an attacker may have some knowledge of the model and expertise to effectively understand the model's data retrieval mechanisms and manipulation of external data by placing malicious prompts (attack complexity is medium (M)). Attack vectors can be external data sources such as websites, APIs, or shared documents that the \ac{LLMs} may access during the data retrieval \cite{greshake2023not}. 

Furthermore, in \textit{prompt divergence}, attackers embed ambiguous or conflicting instructions within the prompt, causing the \ac{LLM} to interpret these instructions in ways that produce divergent responses to deviate from LLM's original goal \cite{wu2024unveiling}. The attacker may understand prompt structure and model processing to create conflicting contexts, which are complex methods requiring high  skills \cite{wu2024unveiling}. Attack vectors can be structured applications in multi-step tasks with diverging instruction encoding techniques.  Therefore, the attack complexity of this attack is high (H).

In \textit{direct prompt injection}, attackers append adversarial commands directly to the system prompt, overriding the LLM's intended functionality \cite{liu2023prompt}. The attacker needs minimal technical knowledge about \ac{LLMs} to append commands \cite{schulhoff2023ignore} and lead to unintended or unsafe outputs. Direct prompt injection attacks are commonly used attack due to its low (L) complexity \cite{schulhoff2023ignore}.

\ 

\noindent\textbf{Embedded Attack}:
In an embedded attack, the attackers craft malicious instructions or manipulate the tokens to carry out harmful actions, e.g., output hazardous knowledge \cite{schwinn2023adversarial} to change the workflow of \ac{LLMs} \cite{schwinn2024soft}. These attacks can be adversarial example attacks or embedded space attacks \cite{wang2022towards}. 

In the \textit{adversarial example attack}, attackers craft small changes to the input that are undetectable for the user but can trick or confuse the model into making erroneous predictions \cite{kumar2024adversarial}. The adversarial example attacker can make small perturbations to the input using basic understanding of input crafting without needing the internal knowledge of \ac{LLMs} \cite{kumar2024adversarial}. Therefore, an adversarial example attack has a low complexity (L).

\textit{Embedded space attacks}, on the other hand, modify the embedding layer of open-source \ac{LLMs} by passing the input string via a tokenised process \cite{wang2022towards}. The user cannot see these modifications since they are hidden in the \ac{LLM}' embedded layers. Attackers utilising embedded space must be proficient in open-source \ac{LLMs} and able to transform input text into hidden token/word representations through specific tactics like gradient descent \cite{wang2022towards}. Thus, embedded space attacks are complex as attackers follow specific tactics and skills in modifying \ac{LLMs}' embedding, resulting in high complexity (H). 

\begin{table*}[!h]
\caption{Attacks on LLMs Model and Attack Complexities }
\label{Complexity}
\begin{tabular}{|l|l|l|l|l|l|}
\hline
\textbf{Attacks} &
  \textbf{\begin{tabular}[c]{@{}l@{}}Attack \\ Sub-types\end{tabular}} &
  \textbf{\begin{tabular}[c]{@{}l@{}}Required \\ Skills\end{tabular}} &
  \textbf{\begin{tabular}[c]{@{}l@{}}Knowledge of Model/\\ Infrastructure\end{tabular}} &
  \textbf{\begin{tabular}[c]{@{}l@{}}Ease of\\ Exploitation\end{tabular}}&
  \textbf{\begin{tabular}[c]{@{}l@{}}Attack \\Complexity\end{tabular}} \\ \hline
\multirow{3}{*}{Multi-Modal} &
  \begin{tabular}[c]{@{}l@{}}Text guided image \\ generation\end{tabular} &  L & L  & L   & \cellcolor{teal}L   \\ \cline{2-6} 
& Cross-modal           & M & L & M & \cellcolor{yellow}M \\ \cline{2-6} 
 & Adversarial image       &  H & M  & H   &\cellcolor{red}H \\ \hline
 \multirow{2}{*}{Flowbreaking}      & Second thoughts             & M & M & M & \cellcolor{yellow}M  \\ \cline{2-6} 
 & Stop and roll           &  M & M  & L   &\cellcolor{yellow}M  \\ \hline
 \multirow{5}{*}{Prompt Injection} & Adversarial prompt      &  M & L  & L   &\cellcolor{teal}L \\ \cline{2-6} 
& Decoy and distract    &  M & M  & L   &  \cellcolor{yellow}M \\ \cline{2-6} 
& Indirect prompting    & M & M & L & \cellcolor{yellow}M \\ \cline{2-6} 
& Prompt divergence     &  H & H  & H   &\cellcolor{red}H \\ \cline{2-6} 
 & Direct injection      &  L & L  & L   & \cellcolor{teal}L \\ \hline
 \multirow{2}{*}{Leakage}      & Data leakage             & M & L & M & \cellcolor{yellow}M  \\ \cline{2-6} 
 & System Prompt leakage           &  H & H  & M   &\cellcolor{red}H  \\ \hline

\multirow{2}{*}{Embedding}        & Embedding space       &  H & H  & H   & \cellcolor{red}H  \\ \cline{2-6} 
& Adversarial example     &  L & L  & L   & \cellcolor{teal}L \\ \hline

\multirow{4}{*}{Jail break}       & Do anything now    &  M & L  & M   & \cellcolor{yellow}M \\ \cline{2-6} 
 & Multi-step jailbreak  & L & M & M  &  \cellcolor{yellow}M \\ \cline{2-6} 
 & Adhoc jailbreak       & L & M & M &  \cellcolor{yellow}M \\ \cline{2-6} 
 & Token-level jailbreak & H & H & H & \cellcolor{red}H \\ \hline
\multirow{4}{*}{Backdoor}         & Dataset backdoor      & L & L & M & \cellcolor{yellow}M\\ \cline{2-6} 
& Pre-trained backdoor  & L & L & L & \cellcolor{teal}L \\ \cline{2-6} 
& Fine-tune backdoor    & H & H & H &  \cellcolor{red}H \\ \cline{2-6} 
& Composite    & H & H & H &  \cellcolor{red}H \\ \hline
 \multirow{3}{*}{Poisoning}    & Dataset poisoning        &  L & L  & L   & \cellcolor{teal}L \\ \cline{2-6} 
 & Pre-trained poisoning   &  L & L  & L   & \cellcolor{teal}L \\ \cline{2-6} 
  & Fine-tune poisoning      & M &  L& M & \cellcolor{yellow}M \\ \hline
\multirow{3}{*}{Inference}    & Membership inference     & M & M & L & \cellcolor{yellow}M \\ \cline{2-6}               & Attribute inference      &  L & L  & L   &\cellcolor{teal}L \\ \cline{2-6} 
 & Input inference         & H & H & H & \cellcolor{red}H\\ \hline
\multirow{2}{*}{Manipulation} & Token manipulation     & H & H & H &  \cellcolor{red}H\\ \cline{2-6} 
 & Content manipulation    &  H & H  &H   & \cellcolor{red}H \\ \hline
\multirow{3}{*}{Extraction}   & Model gradient   & H & H & H &  \cellcolor{red}H \\ \cline{2-6} 
& Training data extraction &  L & L  & L   &  \cellcolor{teal}L \\  \cline{2-6} 
 & Model theft              & H & H & H & \cellcolor{red}H \\ \hline
 \multirow{2}{*}{Evasion}          & Obfuscation           & L  & L & L & \cellcolor{teal}L \\ \cline{2-6} 
 & Token smuggling              &  L & L  & L   & \cellcolor{teal}L \\ \hline
 \multirow{2}{*}{Hallucination}    & Weak semantic               & H & H & H& \cellcolor{red}H  \\ \cline{2-6} 
& Out-of-distribution                  & H  &  H & M  &\cellcolor{red}H \\ \hline

\multirow{2}{*}{Hijacking}  & Square                  &  L & L  & L   &\cellcolor{teal}L   \\ \cline{2-6} & Vocabulary               & H & H & H & \cellcolor{red}H 
 \\ \hline
                              
\end{tabular}
\end{table*}

\ 

\noindent\textbf{Jailbreak Attack:}
Jailbreak attacker bypasses \ac{LLMs}' safety guardrails to respond to unsafe or restricted questions and output inappropriate content such as malware, scams, and illegal or socially harmful instructions \cite{shayegani2023survey,LLM2024,peng2024securing}. Jailbreak attacks come in different forms, such as Do Anything Now (DAN) mode, multi-step prompt, ad-hoc jailbreak and token level jailbreak attacks.

In \textit{DAN mode jailbreak prompt attack}, \ac{LLMs} are being forced to undertake dangerous, harmful actions like ``Do Anything Now" (DAN) mode, preventing them from completing their intended task \cite{gupta2023chatgpt}. This means the attacker prompts the model to act unrestrictedly, effectively using role-playing instructions and unlocking capabilities limited by safety protocols \cite{liu2023autodan}. The DAN jailbreak attacker requires creativity in prompt construction and an understanding of role-play dynamics, instead of a deep technical background. Thus, DAN jailbreak attacker does not need high skills (attack complexity is M) because it uses creative framing to exploit the model's flexibility with user roles, relying more on inventive prompting than in-depth model knowledge. 

In \textit{multi-step jailbreak attacks}, the LLM model's filters are gradually weakened by a series of prompts, eventually reducing the chance that the model will follow its limitations, such as red queen attack \cite{jiang2024red}. Therefore, the attacker does not need specific tools to comprehend how prompts sequencing interacts over a number of steps and how contextual layering can affect model behaviour, which is regarded as medium-level complexity (M) \cite{li2023multi}.

\textit{Ad-hoc jailbreak attacks} are characterised by improvised, creatively crafted prompts to bypass the model’s restrictions \cite{shayegani2023survey}. The attacker directly inputs crafted prompts that manipulate the model into providing restricted information. Techniques include hypothetical scenarios, attention-shifting, and context manipulation, where the prompt creates a scenario or role-play that bypasses ethical guidelines. The ad-hoc jailbreak attacker skill requirement is generally low  because it may mainly involve basic prompt phrasing without deep technical understanding. However, some prompts may require insight into how models interpret instructions, relying on creativity and user insight rather than technical expertise \cite{zhou2024virtual}. Overall, this attack has medium (M) complexity.

Attackers that use \textit{token manipulation jail breaks} take advantage of particular tokens, often anomalous or special text, that the model processes in unusual manners. By employing tokens or sequences that trigger behaviors inconsistent with the model's intended aim, these attacks take advantage of the way models interpret tokenised input \cite{huang2023catastrophic}. Tokenisation has unique effects on model behavior; special tokens like or are frequently used to exploit it. These tokens cause the model to output responses that bypass restrictions, possibly due to the special token's influence in the tokenisation or generation process. This attack requires understanding tokenisation, generation mechanisms, and specific token functions within the LLM's architecture. The attackers require high skill  and technical knowledge  to manipulate tokens to achieve specific behavior within the \ac{LLMs}, making the overall attack complexity high (H) \cite{huang2023catastrophic}.

\ 

\noindent\textbf{Poisoning Attack:}
Poisoning attacks influence the integrity of the training data; attackers can introduce deliberately manipulated data into the model's training phases \cite{bowen2024scaling}. There are different classifications for poisoning attacks, such as pre-training poisoning, dataset poisoning and fine-tuning poisoning.

In \textit{pre-training poisoning}, attackers can inject malicious or biased content into public internet sources such as Wikipedia, a widely used resource for \ac{LLMs} training. This poisoned data may included in the LLM's initial training set. Since \ac{LLMs} rely on massive datasets, minor edits are hard to detect but can substantially affect \ac{LLMs}' behaviour \cite{zhang2024persistent}. Thus, the complexity of this attack is low (L) as the attackers do not need task-specific knowledge or high skills \cite{bowen2024scaling}. 

 In \textit{dataset poisoning}, attackers add harmful or biased content to specific datasets devised for a particular application (e.g., medical sector) domain \cite{datasetpoision}. Thus, dataset poisoning limits selected domains by adding biased or misrepresentative content to carefully selected datasets. This can skew the LLM's behaviour in specific applications, especially where data curation is imperfect, leading to slight but impactful biases in task-specific \ac{LLMs} outputs \cite{bowen2024scaling}. Attackers do not need to manipulate large datasets, but must understand the basic curation process. Thus, knowledge of the dataset's intended application helps attackers insert biases that may not be easily detected in specific domains \cite{zhang2024persistent}. The impact of the dataset poisoning attack is typically limited to specific tasks or fields where curated datasets are applied. Thus, the attack complexity is medium (M).  However, it can create severe biases or misinformation in sensitive applications (e.g., health or legal), affecting user trust in model outputs.  

\textit{Fine-tuning poisoning attack} targets the final training phase, where \ac{LLMs} are fine-tuned for specific tasks or applications. Attackers introduce harmful data to override safety and alignment features, often embedding backdoors that activate with specific input triggers \cite{yao2024poisonprompt}. Attackers exploit fine-tuning APIs to insert carefully crafted triggers or backdoors, which modify the model's behaviour upon receiving specific prompts. This approach allows attackers to bypass moderation controls by embedding hidden behaviours that activate only under specific conditions \cite{zhang2024persistent}. Therefore, fine-tuning poisoning attackers requires specific API knowledge to evade moderation controls effectively and balance subtlety with effectiveness, creating hidden triggers that only activate when intended \cite{wang2022towards}. Overall, the attack complexity is high (H).

\ 

\noindent\textbf{Evasion Attack:}
In an evasion attack, the attacker crafts fake samples during the inference phase, which is not noticeable but leads to incorrect/unexpected behavior \cite{yao2024poisonprompt,xu2020adversarial,vitorino2024adversarial}. Evasion attacks have different forms, such as obfuscation or token smuggling.

In an \textit{obfuscation attack}, an input text is manipulated using word-level or character-level subtle perturbation \cite{vitorino2024adversarial}. The attacker changes the input in several ways, such as by replacing a single word or certain words with similar words, adding special characters, or altering the sentence's structure. Collectively, these techniques make the model confusing, making it difficult for the model to recognize the intended meaning of the input. In this attack, the attacker does not have the authority to change the model's architecture or its parameters. Therefore, to evade the detection, the attacker needs basic obfuscation tactics only, and with trial and error, the attacker could result in harmful, or restricted contents. Overall, the attack complexity of the obfuscation attack is low (L).  

\textit{Token smuggling} attack  comprises the banned words, which are encoded in the attacker's input to evade the filters or detections. The purpose of the attack is to alter the behavior of the model to produce the incorrect output. The complexity of this attack is also low (L) as to launch this attack, the attackers only need to simply split the words and do not need in-depth knowledge of the model.

\ 

\noindent\textbf{Extraction Attack:} In extraction attacks, attackers  use model training data or extract the specific LLM architecture and parameters and recreate the model for execution \cite{birch2023model}. The  attackers can leverage the target LLM by supplying prompts refined to induce the LLM to perform the intended task (e.g., summarisation, chat-based responses, question answering, etc.). This refined prompting process enables attackers to effectively refine and transfer the task-specific capabilities into the extracted model for their purposes \cite{birch2023model}. Extraction attacks could be of several types such as model gradient attacks, training data extraction and model theft.  

In \textit{model gradient attacks}, attackers use precise gradient based training to recreate the model because generally malicious actors cannot steal highly valuable models, such as those trained on rare or hard-to-obtain datasets. This attack poses a significant threat, as it enables the theft of cloud-hosted models without requiring input data. Consequently, such attacks have high (H) complexity as attackers require having an in-depth understanding of the LLM model and infrastructure \cite{miura2024megex}.
 
\textit{Training data extraction} attackers can exploit \ac{LLMs} which are  trained on private datasets. By querying the language model, they can recover individual training samples, extracting verbatim sequences from the model's training data using only black-box query access. This enables attackers to retrieve (publicly available) personally identifiable information (e.g., names, phone numbers, and email addresses) and other non-sensitive information \cite{carlini2021extracting}. Thus, this attack is accomplished in a single step, and the complexity of this attack is classified as low (L).

\textit{Model theft attack} is a black-box adversarial attack. Attackers create an extracted model by deriving specific features (e.g., architecture, parameters, and hyperparameters) from the target model of interest, enabling them to reconstruct it. Once the extracted model is established, attackers can carry out further adversarial attacks, such as model inversion, membership inference, privacy data leakage, and model intellectual property theft \cite{birch2023model}. To execute a model theft attack, attackers require extensive knowledge of \ac{LLMs} to perform several critical steps such as prompt design for crafting prompts to attain task-specific LLM responses, data generation to derive extracting model characteristics, extracted model training for model recreation and ML attack staging against a target LLM \cite{birch2023model}. Due to the complexity and technical depth involved, the complexity of a model theft attack is classified as high (H).

\ 

\noindent\textbf{Backdoor Attack:}
The concept of a backdoor attack is to inject triggers (short phrases, prompts, or instructions) into models, including \ac{LLMs} \cite{yang2024comprehensive}. The attacker inserts triggers into a specific section, such as an open-source library, poisoned training data, etc. \cite{zhao2023prompt,galli2024noisy}. When user inputs are triggered, the model will output some specific contents by the attacker \cite{zhao2023prompt}. The backdoor attacks are of various types, including dataset backdoor, pre-trained backdoor, fine-tuned backdoor, and composite attacks.

In the \textit{dataset backdoor}, attackers deploy poisoned training data in an open-source library. If some LLM developers  utilise it to train their models, they may unknowingly embed a hidden backdoor in the model and open it for manipulation by attackers. This attack doesn't require comprehensive technical skills. The dataset backdoor, however, needs the attacker to perform multi-steps to launch an attack. Hence, the attack complexity of this attack is medium (M) 
\cite{yang2024comprehensive,eykholt2024taking}.
\begin{table*}[!t]
\caption{Attacks on LLMs, Attack Vectors and Impact of Attacks }
\label{vectors}
\begin{tabular}{|p{0.05\columnwidth}|p{0.3\columnwidth}|p{0.8\columnwidth}|p{0.7\columnwidth}|}
% \caption{Attacks on LLMs, Attack Vectors and Impact of Attacks }
\hline
& \textbf{Attacks}              & \textbf{Attack Vectors}                         & \textbf{Impacts }                                                                                             \\ \hline
\multirow{14}{*}{\rotatebox[origin=c]{90}{\begin{tabular}[c]{@{}l@{}}\textbf{LLMs Model-Based Attacks} \end{tabular}}} 
& Multimodal  & Crafted text with an image or sound, adversarial  images,  dataset, poisoned associations (adversarial perturbation), pre-trained model,crafted noise with audio                                & Faulty output, model's operation behavior, misinformation                                                                                    \\ \cline{2-4} 
&Flowbreaking  &    Input prompts, stop button        & Information disclosure, unauthorised action,  incorrect output \\ \cline{2-4} 
&Prompt Injection  &    Input fields, website's code, APIs, encoding techniques, system level-prompts, instruction tuning datasets            & Data breaches, unauthorised action,  erroneous output, misinformation \\ \cline{2-4} 
&Leakage        &    Malicious code, model confidence scores, system prompts                                    & Data breaches, information disclosure                                                                                    \\ \cline{2-4} 
&Embedding   & Crafted instructions, characters or tokens, open source \ac{LLMs} &  Output biases, faulty/erroneous/toxic output  \\ \cline{2-4} 
  % Untargeted  &Model, Input &  Erroneous output/Misprediction  \\ \hline
& Jailbreak        & Heuristic-based prompting, tokens, simple crafted inputs, hypothetical scenarios with acknowledgment, pretrained model        & Tricking LLMs,  incorrect output,  unauthorised access, personal information leakage, privilege escalation, political propaganda               \\ \cline{2-4} 
&Backdoor          & Dataset, open source libraries, triggers, entrusted third party service provider, pre-trained model, trigger keys (one or many)                        & Faulty/incorrect decision, unauthorised access                                                              \\ \cline{2-4} 
&Poisoning  &
  Bad data from unreliable sources,  large amounts of skewed or biased input, publicly available resources, fine-tuned APIs &
  Output biases, unethical behavior, faulty/erroneous results, disinformation,  misinformation \\ \cline{2-4} 
&  Inference           & Dataset, query a particular data-point, network sniffing, compromised APIs& Unauthorised access, data breaches, privacy violation, reputational damage                                                                                         \\ \cline{2-4} 
 &Manipulation  & Tokens, malicious instructions,  fine-tuning process                               & Biased output, customer dissatisfaction, misrepresentation                                                                                              \\ \cline{2-4} 
 & Evasion  &Fake samples, simple, banned, similar words, special characters&   Data breaches, Incorrect output \\ \cline{2-4} 
& Extraction &
  Datasets, gradients queries, spoofing by trusted parties, biometric, crafted code, open API, model architecture and parameters, poorly configured outputs  & Incorrect/unexpected predictions, biased output, information disclosure, financial losses
   \\ \cline{2-4} 
&Hallucination       &    Fabricated code libraries, semantic input craft, random tokens                        &     Nonsensical/unfaithful output                                                                      \\ \cline{2-4} 
& Hijacking attack       &    Delimiters or instructions, randomised search methods, LLM vocabulary                                    & Information disclosure, false output, unauthorised control, offensive behavior                                                                         \\  \hline
% Adversarial Example & Input                                  & Misinformation                                                                                              \\ \hline
% Sabotage            & Model's normal operation               & Operational disruptions \\ \hline

% Espionage           & Data logs, Models                      & Data breach, Unauthorised access   
\multirow{8}{*}{\rotatebox[origin=c]{90}{\begin{tabular}[c]{@{}l@{}}\textbf{LLMs Infrastructure-Based Attacks} \end{tabular}}} 
&  Unbounded consumption  & Numerous crafted inputs, complex and resource-intensive queries, cloud-based AI services & Model service degradation, financial losses, reputational damage, service unavailability, system failure   \\ \cline{2-4} 
& Inadequate Sandboxing & Passwords, API keys, files/network access, plugins’ permissions, misconfigurations & Unauthorised access and action, cross-system exploitation,  privilege escalation, data corruption/loss.    \\ \cline{2-4} 
& Access Control  &Access control policies, API, file/ network information, social engineering tactics, by-default configurations, arbitrary codes execution on server &  Data breaches, privilege escalation, misinformation, harmful output, public distrust \\ \cline{2-4} 
& Supply Chain        &          Datasets, compromised or outdated libraries or models, pre-trained models, insecure plugins or APIs, misconfigurations, unclear policies or agreements  & Service outage, privilege escalation, unauthorised action, data breach, biased output, network disruption\\ \cline{2-4}
&   Side-Channel &Model parameters and architecture information, training data, response time, API, power consumption information&   Information disclosure, system exploitation \\ \cline{2-4} 
  &  Server-side Request Forgery & \begin{tabular}[c]{@{}l@{}}Inputs, security misconfiguration, internal services \\  access request, API, secured data stores \end{tabular} & \begin{tabular}[c]{@{}l@{}} Unauthorised access, model malfunctioning, \\data exfiltration, Tempering \end{tabular} \\ \cline{2-4}
&  Remote Code Execution  & Code, shell commands, arbitrary code execution& Network disruption, system unavailable,  unauthorised access  \\ \cline{2-4}
& Ransomware  &Publicly available code repositories, databases, CVEs, fine-tuned models &  System unavailable, data breach reputational damage, financial losses  \\ \hline
   % Model Theft/Stealing \cite{kariyappa2021maze} & Model's parameter &  \begin{tabular}[c]{@{}l@{}}Unauthorised access, economic losses, information \\ disclosure, copying or exfiltration of models data \end{tabular} \\ \hline
   % Model Replication  &&    \\ \hline
 
  %  SQL Injection Attack \cite{pedro2023prompt,schulhoff2023ignore} &Input&  Tricking LLMs, malicious action generation  \\ \hline
%  Goal Hijacking  \cite{schulhoff2023ignore, levi2024vocabulary} &Input&  Harmful Information generation  \\ \hline
%   Token Wasting \cite{schulhoff2023ignore} &Tokens &   Costs on the application’s
% maintainer  \\ \hline
    
\end{tabular}
 
\end{table*}

In the \textit{pre-trained attacks}, the attacker is assumed to be an untrusted third-party service provider and offers (or open-source) pre-trained \ac{LLMs}, which are tailored to specific targets (such as datasets or prompt templates) designed to attract potential users. Consequently, the attacker has complete control over the training dataset and the training process of the target model, so attacker does not require internal knowledge of the model and does not require specific expertise to launch an attack (attack complexity is low (L)) \cite{galli2024noisy}.

The \textit{fine-tuned backdoor attack} is similar to a dataset backdoor, but attackers need more technical skills. For example, attackers embed a backdoor into a model and upload it to the Internet, waiting for some unsuspecting victims (like developers) to download this model. The difference is when the developer proceeds with models to fine-tune them for a specific purpose, the attacker operates in a white-box environment and modifies the model’s parameters, structure, and training data \cite{yang2024comprehensive}. These vectors require the attacker to be very familiar with the trained model and skilled in model training. Thus, fine-tuning backdoor attack needs in-depth knowledge and highly complex skills for launching an attack on the \ac{LLMs}. Overall, sophistication level of this attack is medium (M).

For the \textit{composite backdoor attack} (CBA), attackers need to set up triggers just like any other backdoor attack, but in this attack, multiple trigger keys are scattered across different prompt components \cite{huang2023composite}. The composite backdoor will only be activated when all trigger keys coincide. The composite backdoor attack is considered more complex and requires advanced expertise, and the attacker must understand the model’s internal workings to execute it effectively \cite{zhao2023prompt}. Therefore, the attack complexity level is high (H) as composite backdoor attacks require high skill and technical knowledge.

\ 

\noindent\textbf{Inference Attack:}
In inference attacks the attacker's motive is to illegitimately retrieve the victim's sensitive information from the \ac{LLMs} \cite{hu2022membership}. \ac{LLMs} tend to memorise information from their training data, and attackers investigate that memorisation of training data. There are three main sub-types of attacks on \ac{LLMs}: membership inference attacks (MIA), attribute inference (model inversion), and input inference attacks. 

\textit{Membership inference attack} (MIA) is one of the most basic forms of inference attack, which allows attackers to fetch data to determine whether a given sample belongs to a training dataset. The attacker’s goal is to determine if a specific data point was used in the training dataset of LLMs by analysing its output, such as memorisation of training data, copyright violations, and test-set contamination \cite{galli2024noisy}. Attackers do not require deep technical background for this attack but only deal with the training dataset and data points. As a result, this is a medium complexity (M) attack. 

\textit{Attribute inference attack} can extract various characteristics of the victims, like ethnicity or gender information from a model, even if this information was not explicitly included in the training data \cite{jayaraman2022attribute,zhao2021feasibility}. In this attack, attackers only need to utilise some simple steps to fetch private information. Therefore, the complexity
of an attribute inference attack is low (L).

In an \textit{input inference attack}, attackers may need a way to intercept the user input to \ac{LLMs}, using other methods such as network sniffing, exploiting compromised APIs, backdoor attacks, or combining these techniques in a composite attack strategy. Only after obtaining the user’s input data can the attacker carry out an inference attack on sensitive information. Consequently, this type of attack requires attackers to have highly advanced technical skills due to its composite nature, so the complexity of input inference is high (H). 

\ 

\noindent\textbf{Manipulation Attack:} 
\ac{LLMs}  may be vulnerable to potential manipulation attacks, which results into public distrust, reputational damage or biased output or misrepresentation \cite{kaghazgaran2019wide}. These attacks allow an attacker to manipulate the  model's generated output, enabling malicious samples to evade
detection without affecting the overall system performance. Attackers can utilise trusted data sources to inject malicious content so that they can introduce manipulated data into the training pipeline by compromising the data source or intercepting it in transit \cite{liao2018server}. The manipulation attacks include two sub-types of attacks, token manipulation and content
manipulation. Please note: token-level jailbreak attack could also lie in this category.

In a \textit{token manipulation} attack, the attackers exploit the vulnerabilities in the process of tokenisation through token substitutions, removals, and syntactic reordering to generate incorrect outputs\cite{gereti2024token}. The attacker needs in-depth knowledge of natural language processing tasks (tokenisation process), and detail knowledge of the model's internal architecture and parameters. This attack could also lie under the category of token level jailbreak. The overall complexity of the attack is high (H). 

In a \textit{content manipulation attack}, attackers can manipulate models to generate fake content and spread AI-generated fake news (disinformation) and social bots on social media platforms. They may also use \ac{LLMs} to produce targeted user outputs, deceiving the public for profit. This attack is relatively easy to execute once attackers gain control of \ac{LLMs} \cite{cresci2021adversarial,alla2025cyberattacks}. Therefore, attackers use specialised methods to craft the ambiguous input and need good understanding of the model behavior to achieve their goals. The complexity of content
manipulation attack is classified as high (H).

 \

\noindent\textbf{Leakage Attack:} In the leakage attack, \ac{LLMs} accidentally leak sensitive and confidential information from their training data through the responses. These attacks include two sub-types, data leakage and prompt leakage. 

In \textit{data leakage attack}, attackers exploit the model's memorisation of sensitive training data to infer, extract, or misuse private information. Such attacks leverage techniques like membership inference or data extraction to recover portions of the training data, potentially including personally identifiable information (PII) or other confidential content, from the model's outputs \cite{lukas2023analyzing}. Attackers need advanced expertise, understanding, and skills to launch data leakage attacks on the \ac{LLMs}. Overall, the attack's sophistication is medium (M).

\textit{System prompt leakage attack} is a specialised attack targeting \ac{LLMs}, which, if uncovered, facilitates other types of attacks \cite{LLMsecurity2025}. Since the functionality and performance of \ac{LLM} applications heavily rely on the system prompt, which directs the underlying \ac{LLMs} on what tasks to perform, developers typically keep these system prompts confidential. In this attack, an attacker sends instructions to the target LLM application, and its responses inadvertently reveal the system prompt (such as information describing various roles and permissions, connection strings, or passwords). Unlike jailbreak attacks, the ultimate goal of prompt leakage is to replicate the same, precise system prompt \cite{hui2024pleak}. The prompt leakage attack requires considerable skill to understand the model architecture and prompt engineering, and the attacker can accomplish its motive by trying different instructions within the \ac{LLMs}, making the attack complexity high (H).

\noindent\textbf{Hijacking Attack:} In a hijacking attack, attackers use controlled instructions to take unauthorized control or exploit the behavior, output, or functionality of an \ac{LLM} for malicious purposes \cite{yao2024poisonprompt,qiang2023hijacking}. There are two types of hijacking attacks, vocabulary and square.

In a \textit{vocabulary attack}, attackers manipulate \ac{LLMs} by inserting delimiters or systematically rephrasing instructions until they achieve their goal—such as revealing confidential information, generating specific false information, or exhibiting offensive behavior \cite{levi2024vocabulary}. To execute this, attackers first identify words in the LLM's vocabulary that trigger the desired target behavior when included anywhere in the user prompt. These words are referred to as adversarial vocabulary \cite{qiang2024hijacking}. This attack is typically the hardest to detect in user prompts using filters or other pattern-matching defenses, as many system prompts are designed to ensure a certain level of robustness in LLM applications, and some \ac{LLMs} include automatic text filters for detection. However, the attack requires attackers to understand the specific model in-depth and to optimise arbitrary word sequences inserted into prompts to alter the model's behavior \cite{levi2024vocabulary}. Therefore, the skill complexity required for this attack is classified as high (H).

\textit{The square attack} is based on a  randomised search strategy that involves selecting square-shaped localized updates at random positions of the input text \cite{qiang2023hijacking}. This ensures that, in each iteration, the perturbation remains roughly on the boundary of the feasible set of the input text \cite{andriushchenko2020square}. The square attack requires minimal expertise in LLMs, making its complexity low (L) to manipulate the input and produce the incorrect output. 

\ 

\noindent\textbf{Hallucination Attack}:  Hallucination attack is possible due to the nature of \ac{LLMs} where attackers reveal the vulnerability in \ac{LLMs} during the inference phase and manipulate them to generate fabricated outputs when users are querying the model \cite{yao2023llm}. Poor benign prompt engineering or just badly functioning models cause these attacks, which results in \textit{excessive agency} vulnerability \cite{LLMsecurity2025}. Thus, attack vectors for such attacks are fabricated code libraries. There are different classifications for hallucination attacks, such as weak semantic attacks and out-of-distribution (OoD) attacks. 

\textit{Weak semantic attacks} are the attacks in which attackers alter a small number of tokens with semantic input and trick the model into generating false information \cite{wu2024unveiling}. The attackers uses a gradient-based token replacement approach for the replacement of few token and insert a perturbed prompt instead for maintaining the semantic of the input. This attacks need in-depth knowledge of the model and advances techniques for understanding the semantics of programming language used, so the attack complexity is high (H).  

The \textit{out-of-distribution (OoD) attacks} use random tokens (semantics are not preserved) that do not match with training data relevant to \ac{LLMs}. As a result, \ac{LLMs} fabricate non-sensible or unfaithful outputs \cite{yao2023llm}. The attacker needs to understand the distribution of training data, prompt engineering, decoding strategies and in-depth knowledge of the working of the model, therefore complexity of the attack is high (H). 
     
\subsection{Security Attacks on LLMs Infrastructure} \label{sec:taxinfr}

The methodologies, various vectors, and impacts of security attacks on the \ac{LLMs} infrastructure (Figure~\ref{fig:infr}) are presented in this section. The analysis of the complexities of each identified attack is elaborated in Table~\ref{Infrastructure}. 

% Table~\ref{Infrastructure} shows the attack complexities of various attacks on \ac{LLMs} infrastructure .

  \

\noindent\textbf{Supply Chain Attack:}
The \ac{LLMs} supply chain involves the whole lifecycle, from model training to ongoing maintenance.
Supply chain attacks infiltrate various stages of the \ac{LLMs} infrastructure, including data preparation, data pre-processing,  model training, model deployment, model optimisation, etc. and exploit vulnerabilities in the components of each stage \cite{LLMsecurity2025,singla2023empirical}. The attacker may inject poisoned data into training, alter model during training or deployment, upload compromised  models to public repositories, or manipulate third-party libraries or code, that support \ac{LLMs} development. It may also involve the exploitation of insecure APIs or LLM plugin extensions that cloud providers use to host \ac{LLM} infrastructure and target deprecated model dependencies or terms and conditions, and copyright material \cite{hu2024large,supplychain}. \ac{LLM} supply chain attack results into data breaches, output manipulation and denial of services as well. Depending on the type of supply-chain attack, the attack complexity could vary. From the perspective of infrastructure level attack, attacker needs in-depth understanding of \ac{LLMs} and needs to have sophisticated technical skills, therefore, the overall attack complexity is high (H). 
\begin{table*}[!ht]
\caption{Attacks on  LLMs Infrastructure and Attack Complexities }
\label{Infrastructure}
\begin{tabular}{|l|l|l|l|l|l|}
\hline
\textbf{Attacks} &
  \textbf{\begin{tabular}[c]{@{}l@{}}Attack \\Sub-types \end{tabular}}&
  \textbf{\begin{tabular}[c]{@{}l@{}}Required \\Skills\end{tabular}} &
  \textbf{\begin{tabular}[c]{@{}l@{}}Knowledge of Model/\\ Infrastructure\end{tabular}} &
  \textbf{\begin{tabular}[c]{@{}l@{}}Ease of \\Exploitation\end{tabular}} &
  \textbf{\begin{tabular}[c]{@{}l@{}}Attack \\ Complexity\end{tabular} }\\ \hline
  \multirow{2}{*}{\begin{tabular}[c]{@{}l@{}}Unbounded \\ Consumption\end{tabular}} & Denial of service
   & M
   & M
   & M
   &
  \cellcolor{yellow}M  \\ \cline{2-6} 
 & Denial of wallet
   &
   L&
   M&
  M &
  \cellcolor{yellow}M  \\ \hline
\multirow{2}{*}{\begin{tabular}[c]{@{}l@{}}Inadequate \\ Sandboxing \end{tabular}} &
    \begin{tabular}[c]{@{}l@{}}Environmental \\ segregation\end{tabular} &
  H &
  H &
  H &
  \cellcolor{red}H  \\ \cline{2-6} 
 &
  System-level interaction &
  H &
  H &
  H &
  \cellcolor{red}H  \\ \hline
  Access control & 
   & H
   & H
   & M&\cellcolor{red}H 
   \\ \hline 
  Supply Chain Attack &
   & H
   & H
   & H
   & \cellcolor{red}H 
   \\ \hline
  \multirow{2}{*}{Side-Channel Attack} & Timing attack
   & L
   & L
   & L
   &
  \cellcolor{teal}L  \\ \cline{2-6} 
 & Power analysis
   &
   M&
   H&
  H &
  \cellcolor{red}H  \\ \hline
\begin{tabular}[c]{@{}l@{}}Server-side Request\\ Forgery\end{tabular} &
   &
  H &
  H &
 H  &
  \cellcolor{red}H  \\ \hline
\begin{tabular}[c]{@{}l@{}}Remote Code \\ Execution \end{tabular} &
   & M
   & M
   & M
   & \cellcolor{yellow}M 
   \\ \hline
Ransomware Attack 
   & 
   & H
   & H
   & H
   & \cellcolor{red}H\\ \hline
   
\end{tabular}
\end{table*}

\ 
     
\noindent\textbf{Inadequate Sandboxing:}
Sandbox attacks exploit the isolated environment (sandbox) where \ac{LLMs} run to execute unintended commands, access unauthorised information, or manipulate the model’s behaviour \cite{abdali2024securing}. These attacks could be of two types: environmental segregation and system-level interaction. 

Attackers target vulnerabilities within the sandbox environment and compromise the LLM’s interaction with external components such as operating system, other containers or virtual machines, resulting into unauthorised access and information disclosure \cite{wu2024new}. Such an attack is called \textit{environmental segregation} attack.  The attack complexity is high (H) as attacker needs high skills and technical knowledge for sandbox operations, isolation handling, and an exploit development environment \cite{abdali2024securing}. 

\textit{System level interaction attack}, due to inadequate sandboxing, exploit the direct interaction of \ac{LLMs} with system-level processes, APIs, hardware components, and shell commands and  result in privilege escalation, and unauthorised access, making the overall sophistication level of an attack high (H).

 \

\noindent\textbf{Access Control Attack} is the exploitation of vulnerabilities in the access control policies or mechanisms that restrict and manage unauthorised access and export to the model’s output, data, and parameters. The attacker uses vectors such as API access, file or network information, arbitrary code execution on servers, social engineering tactics, or by-default configurations, resulting in privilege escalation, data breaches, misinformation, or harmful output. The attacker needs to know the API, network configurations, authentication, and authorisation mechanisms, or have expertise in prompt engineering. Overall, the access control attack complexity is high (H).  

\ 

\noindent\textbf{Ransomware Attack:}
Ransomware attackers target the model infrastructure or data to compromise \ac{LLMs} operations. Attackers can encrypt, lock, or manipulate the LLM’s functionality to disrupt its usability or extract sensitive information, which cannot be used until the ransom is paid \cite{zhang2024llms}. The attacker can exploit training datasets, pre-trained models, explore CVEs to identify unpatched vulnerabilities, and third-party code to launch an attack. Therefore, attackers require advanced knowledge of the LLM architecture, APIs, hosting environment, and cryptographic methods and need high skills in AI-specific deployments \cite{itonin2024leveraging,zhang2024llms}. The attack complexity of a ransomware attack is high (H).  

 \ 

\noindent\textbf{Unbounded Consumption Attack:}
An unbounded consumption attack is a malicious attempt in which an attacker makes the \ac{LLMs} services unavailable for legitimate users or involved in the target's financial resources depletion. To launch unbounded consumption attack, the attacker exploits the LLM's ability to generate uncontrolled responses based on input queries, resulting in resource exhaustion, system failure, and economic losses  \cite{LLMsecurity2025}. Unbounded consumption attacks are categorised as denial-of-service (DoS) (resource-exhaustion attack) and denial of wallet (DoW) attack.  

In \textit{denial-of-service or resource-exhaustion attack}, attackers craft inputs to disrupt or degrade the services of \ac{LLMs} for legitimate users. The attacker overwhelms \ac{LLMs} with varying lengths of inputs, sends them the sheer volume of inputs that exceed the LLM's context window, or submits complex or resource-intensive queries to perform resource-heavy operations. Resource-exhaustion attacks result in resource depletion, increased latency, degraded performance, or even complete service unavailability, unresponsiveness, and potential failures. The attacker should have knowledge of prompt engineering or can use tools to send a large number of prompts, needs knowledge of the model, such as the token or context window limit, and infrastructure, so the attack complexity is medium (M). 

In \textit{denial of wallet (DoW) attack}, attackers exploit the financial model of cloud-based AI services by performing a high volume of operations, impacting  financial sustainability of the service provider. The attacker could use bots, have knowledge of prices of the services and have some understanding of model and cloud service provider security measures to bypass them. These factors make the overall attack complexity medium (M).

\

\noindent\textbf{Side-Channel Attack:} Side-channel attacks exploit the model parameters and architecture information, system's physical or logical operations information to infer sensitive user data \cite{LLMsecurity2025}.  Attackers utilise training data filtering, input preprocessing, and query filtering against language models to cause data leakage in models 
\cite{debenedetti2024privacy,kulkarni2023order}. 

The attackers could perform \textit{side-channel timing attacks} where attacker could use tools to determine the response times of the queries  \cite{zheng2024inputsnatch}. The attacker needs the basic knowledge of timing variations and does not need to understand the internal architecture of the model or the system; only basic information about the API to measure response time is required. Timing attack has overall low (L) complexity. 

A \textit{power analysis side-channel attack} on \ac{LLMs} is launched to acquire sensitive information about the data, parameters, and architecture of \ac{LLMs} by exploiting variations in the power consumption of the LLM's hardware. The attack complexity is high (H) as attackers need in-depth knowledge of the underlying infrastructure, hardware configuration, and the identification of meaningful insights from noisy data and shared memory systems \cite{nazari2024llm}. 

\ 

\noindent\textbf{Server-side Request Forgery:} In this attack, the attackers target vulnerabilities in the servers where the \ac{LLMs} are deployed and exploit weaknesses in token transmission, API endpoints, or the network infrastructure to intercept, manipulate, or extract sensitive data. Therefore, attackers need high skills to monitor encrypted traffic and analyse token-length sequences to reconstruct \ac{LLMs}' responses \cite{weiss2024your}. Overall, the attack complexity of this attack is high (H).

\ 

\noindent\textbf{Remote Code Execution Attack:}
Remote Code Execution (RCE) attacks are infrastructure attacks where attackers exploit vulnerabilities in software systems to execute arbitrary code on the target machine.  These attacks are due to improper output handling by \ac{LLMs} \cite{LLMsecurity2025}. The attacker provides prompts such as code, shell commands, or specific operations executable by the shell or interpreter, and the motive of an attack is to manipulate the \ac{LLMs} to produce dangerous output. 
The attack needs prompt-engineering knowledge, knowledge of shell commands, and the knowledge of \ac{LLMs} operations and interactions  \cite{liu2023demystifying}. These attacks lead to unauthorised access, data disclosure, and system compromise \cite{liu2023demystifying}. Attackers use multiple steps to manipulate the model’s behavior in this attack. Therefore, this is of a medium (M) complexity attack.

\section{Application of Taxonomy on Education Sector with Dread Model }
\label{sec:application}
This section presents various categories of the DREAD threat model. We also present the rationale for the selection of scores/levels for each of the categories. Then, the proposed taxonomy is mapped on the education sector using the DREAD score.

\subsection{DREAD Model Categories, Scores and Rationale } \label{dreadmodel}

DREAD provides a structured and quantitative approach to assess and prioritise security threats based on a risk score, calculated using five criteria.  These criteria are:  (1) Damage (impact of an attack), (2) Reproducibility (ease of reproducing/replicating an attack), (3) Exploitability (the effort required to launch an attack), (4) Affected Users (number of (end) users affected by a threat being exploited), and (5) Discoverability (likelihood of a threat being exploited (discovered)). A threat receives a score of 0 to 10 for each category. The final rating of the threat is calculated based on the individual scores, and then the average score (overall risk score) is taken. Please note: the risk score ratings for each category in this study are based on subjective observations by security and educational experts in our team and  would likely vary when assessed by other experts. 

 \

\noindent{\textbf{Damage}:} In the educational sector, the attacks on \ac{LLMs} can damage or have an impact on student, researcher, employee (faculty members and professional staff) personal or financial information, academic integrity, or research data (intellectual property) or other institutional data, institution's infrastructure, including networks, applications, and devices, or involve reputational damage, institutional financial losses or operational discontinuity. Table~\ref{damagescore} shows the damage scores and their respective rationales. 

\begin{table}[!h]
\caption{Damage score and rationales}
\label{damagescore}
\begin{tabular}{|l|p{0.8\columnwidth}|l|}
\hline
\textbf{Score} & Rationale      \\ \hline
0     & No Damage      \\ \hline
2.5            & Non-sensitive data exposure \\ \hline
5   &  Output biases/ unethical behavior/ faulty/erroneous/misleading output                         \\ \hline
7.5 & Privilege escalation/ misprediction/ personal information disclosure/ data breach \\ \hline
10  & Operational disruption/ financial losses/  disinformation (reputational damage)             \\ \hline
\end{tabular}
\end{table}

\ 

\noindent{\textbf{Reproducibility};} The reproducibility of \ac{LLM}-based attacks can range from easy to circumstantial (Table~\ref{reproducescore}). This categorisation is based on the number of steps  or internal details of the model. 

% -Easy: Using simple or single step, or does not require the internal knowledge of the model or the infrastructure. 

% -Complex: Using multi-step, but does not require the internal knowledge of the model or the infrastructure.

% -Very Complex: use of multi-steps and detail internal knowledge about the model is required

% -Impossible: Extremely difficult or impossible to reproduce the attack. 

\begin{table}[!h]
\caption{Reproducibility score and rationales}
\label{reproducescore}
\begin{tabular}{|l|l|p{0.5\columnwidth}|l|}
\hline
\textbf{Score} & Meaning   & Rationale   \\ \hline
0    & NA    &  \\ \hline
2.5  & Circumstantial    &  Extremely difficult or impossible to reproduce the attack.   \\ \hline
5 & Very Complex & Use of multiple steps and in-depth internal knowledge about the model is required. \\ \hline
7.5   & Complex  &    Using multiple steps, but does not require the internal knowledge of the model or the infrastructure. \\ \hline
10   & Easy & Using a single step, or does not require the internal knowledge of the model or the infrastructure. \\ \hline
                     
\end{tabular}
\end{table}
\ 

\noindent{\textbf{Exploitability:}} The exploitability of an attack could be determined from the skills or experience required by the attacker. For this category, the relevant scores are mentioned in Table~\ref{exploitablescore} and rationale for these scores are as follow:

% -Easy: Attacker use available tools or does not require any skills or expertise.

% -Complex: attacker needs some skills or experience with some sophisticated techniques and available tools.

% -Very Complex: use of  specialised or advance tactics, skills, methods or tools.

% -Impossible: Difficult or impossible to launch an attack.

\begin{table}[!h]
\caption{Exploitability score and rationales}
\label{exploitablescore}
\begin{tabular}{|l|l|p{0.5\columnwidth}|l|}
\hline
\textbf{Score} & Meaning &Rationale      \\ \hline
0    & NA  &    \\ \hline
2.5  & Circumstantial  &   Extremely difficult or impossible to launch an attack. \\ \hline 
5 & Very Complex & Use of  specialised or advance tactics, skills, methods or tools.\\ \hline
7.5   & Complex   &  Attacker needs some skills or experience with some sophisticated techniques and available tools.                    \\ \hline
10    & Easy & Attacker uses available tools, publicly available information or does not require any skills or expertise.\\ \hline
\end{tabular}
\end{table}

\ 

\noindent{\textbf{Affected Users:}} There are a number of internal or external stakeholders that could be impacted or affected by an attack on \ac{LLMs} in an educational institution. The affected users could be students, operational staff, administrative staff, academic staff, personnel from upper management, board members, etc. The attack could impact an individual such as, a single student or staff member, or a group of people, such as a research team (students, researchers, and staff) or an administrative team such from admissions.

%Note: This category is very important in assessing the risk involved with a particular attack. To provide a holistic assessment, we can sub-categorise and integrate this category with the Damage category. However, we have selected the higher score for this category (intentionally) because with an attack, all the stakeholders get impacted directly or indirectly.  Table~\ref{userscore} shows the  scores and their respective rationales.

This category is similar to the Damage category, however, here we only consider the number and type of users affected by the attack rather than overall damage. Generally, an attack is scored higher if more users are affected from the attack, and vice versa. We have classified attacks affecting Admin users or higher management as 7.5 since such attacks potentially affect a larger number of end-users. Where it is felt that the attack may be conducted in ways that affect different number of users, we choose the highest possible score. 
\begin{table}[!h]
\caption{Affected User score and rationales}
\label{userscore}
\begin{tabular}{|l|p{0.8\columnwidth}|l|}
\hline
\textbf{Score} & Rationale      \\ \hline
0    & No User(s)      \\ \hline
2.5            & An individual user- student/ staff/researcher \\ \hline
5   & Group of users                        \\ \hline
7.5 & Administrative user(s) or higher management individual(s) \\ \hline
10  & All stakeholders             \\ \hline
\end{tabular}
\end{table}

\ 

\noindent{\textbf{Discoverability}:} The ease with which an attack is able to discover a vulnerability depends on the attack vectors utilised by the attacker. The level of the discoverability could be easy, complex, very complex and impossible as shown in the Table~\ref{discover}. The rationale for the each level is as follows:

% -Easy: Discovery based on simple input prompts or publicly available information

% -Complex: Discovery based on crafted instructions

% -Very Complex: use of heuristic based prompting

% -Impossible: Difficult or impossible to discover a vulnerability.

\begin{table}[!h]
\caption{Discoverability score and rationales}
\label{discover}
\begin{tabular}{|l|l|p{0.5\columnwidth}|l|}
\hline
\textbf{Score} & Meaning &Rationale      \\ \hline
0     &   NA  &  \\ \hline
2.5  &   Circumstantial  &   Difficult or impossible to discover a vulnerability \\ \hline
5 &  Very Complex & Use of heuristic based prompting or in-depth knowledge of the model or infrastructure is required \\ \hline
7.5   &          Complex &   Discovery based on the crafted instructions or or some knowledge of the model or infrastructure is required         \\ \hline
10           &  Easy & Discovery based on simple input prompts or trial or error basis, or internal details of the model or infrastructure not required\\ \hline

\end{tabular}
\end{table}

\begin{table*}[!h]
\caption{Risk Assessments of the Attacks on LLMs  in Education Sector using DREAD (Damage, Reproducibility, Exploitability, Affected Users, Discoverability) Model }
\label{educationtable}
% \scriptsize
\begin{tabular}{|l|p{0.3\columnwidth}|l|p{0.1\columnwidth}|p{0.1\columnwidth}|l|l|p{0.1\columnwidth}|l|l|l|p{0.1\columnwidth}|l|p{0.1\columnwidth}|l|l|}
\hline
&\textbf{Attacks} &
  \textbf{Attack Sub-Types} &
  \textbf{D} &
  \textbf{R} &
  \textbf{E} &
  \textbf{A} & \textbf{D} & \textbf{Risk Score} & \textbf{Level}\\ \hline
\multirow{13}{*}{\rotatebox[origin=c]{90}{\begin{tabular}[c]{@{}l@{}}\textbf{LLMs Model-Based Attacks} \end{tabular}}}&\multirow{3}{*}{Multi-Modal} &
  \begin{tabular}[c]{@{}l@{}}Text guided image \\ generation\end{tabular} &5&10&10&7.5&10&8.5&\cellcolor{orange}H \\ \cline{3-10} 
&& Cross-modal           &5&7.5&7.5&7.5&10&7.5& \cellcolor{orange}H\\ \cline{3-10} 
& & Adversarial image       &7.5&5&5&7.5&7.5&6.5& \cellcolor{yellow}M \\ \cline{2-10}
&\multirow{2}{*}{Flowbreaking}        & Second thoughts       &7.5& 5&7.5& 2.5 &7.5& 6 & \cellcolor{yellow}M\\ \cline{3-10} 
&& Stop and roll  &7.5&7.5&10&2.5&10&7.5& \cellcolor{orange}H\\ \cline{2-10}
&\multirow{5}{*}{Prompt Injection} & Adversarial prompt      &7.5&10&10&7.5&10&9&\cellcolor{red}C \\ \cline{3-10} 
&& Decoy and distract    &5&7.5&10&5&7.5&7&\cellcolor{orange}H \\ \cline{3-10} 
&& Indirect prompting    &7.5&7.5&7.5&10&7.5&8&\cellcolor{orange}H\\ \cline{3-10} 
&& Prompt divergence     &7.5&5&5&5&5&5.5&\cellcolor{yellow}M\\ \cline{3-10} 
& & Direct injection      &5&10&10&10&10&9&\cellcolor{red}C \\ \cline{2-10}
&\multirow{2}{*}{Embedding}        & Embedding space       &5& 7.5&5& 7.5 &5& 6 & \cellcolor{yellow}M\\ \cline{3-10} 
&& Adversarial example  &5&10&10&7.5&10&8.5& \cellcolor{orange}H\\ \cline{2-10}
 &\multirow{2}{*}{Evasion}          & Obfuscation           &5&10&10&7.5&10&8.5&\cellcolor{orange}H\\ \cline{3-10} 
& & Token smuggling              &7.5&10&10&7.5&10&9&\cellcolor{red}C\\ \cline{2-10}
&\multirow{4}{*}{Jail break}       & Do anything now    &5&7.5&7.5&10&10&8&\cellcolor{orange}H \\ \cline{3-10} 
 && Multi-step jailbreak  &10&10&7.5&10&7.5&9& \cellcolor{red}C\\ \cline{3-10} 
 && Adhoc jailbreak       &7.5&10&7.5&7.5&7.5&8& \cellcolor{orange}H\\ \cline{3-10} 
 && Token-level jailbreak &5&5&5&7.5&5&5.5& \cellcolor{yellow}M\\ \cline{2-10}
 & \multirow{3}{*}{Poisoning}    & Dataset poisoning        &5&10&10&5&10&8&\cellcolor{orange}H\\ \cline{3-10} 
 && Pre-training poisoning   &5&10&10&5&5&7& \cellcolor{orange}H\\ \cline{3-10} 
 & & Fine-tuned poisoning     &7.5&7.5&7.5&7.5&10&8&\cellcolor{orange}H \\ \cline{2-10}
&\multirow{4}{*}{Backdoor}         & Dataset backdoor      &5&10&7.5&5&10&7.5&\cellcolor{orange}H\\ \cline{3-10} 
&& Pre-trained backdoor  &5&10&7.5&5&10&7.5&\cellcolor{orange}H\\ \cline{3-10} 
&& Fine-tuned backdoor    &5&5&10&5&10&7&\cellcolor{orange}H\\ \cline{3-10} 
&& Composite    &7.5&5&5&5&5&5.5& \cellcolor{yellow}M \\ \cline{2-10} 
&\multirow{3}{*}{Inference}    & Membership inference     &7.5&7.5&10&7.5&10&8.5& \cellcolor{orange}H\\ \cline{3-10}               & &Attribute inference      &2.5&10&10&10&10&8.5& \cellcolor{orange}H\\ \cline{3-10} 
& & Input inference         &7.5&5&5&10&5&6.5& \cellcolor{yellow}M\\ \cline{2-10}
&\multirow{2}{*}{Manipulation} & Token manipulation     &5&5&5&7.5&10&6.5&\cellcolor{yellow}M\\ \cline{3-10} 
 && Content manipulation    &10&5&5&10&5&7& \cellcolor{orange}H\\ \cline{2-10} 
&\multirow{3}{*}{Extraction}   & Model gradient   &5&5&5&10&5&6& \cellcolor{yellow}M\\ \cline{3-10} 
&& Training data extraction &7.5&10&10&7.5&7.5&8.5& \cellcolor{orange}H\\  \cline{3-10} 
& & Model theft              &10&5&5&10&5&7&\cellcolor{orange}H\\ \cline{2-10} 
&\multirow{2}{*}{Leakage}      & Data leakage             &7.5&7.5&7.5&10&10&8.5&\cellcolor{orange}H  \\ \cline{3-10} 
& & System prompt leakage           &7.5&5&7.5&10&5&7& \cellcolor{orange}H \\ \cline{2-10} 
& \multirow{2}{*}{Hallucination}    & Weak semantic               &7.5&5&5&7.5&5&6& \cellcolor{yellow}M \\ \cline{3-10} 
&& Out-of-distribution                  &7.5&5&5&5&5&5.5& \cellcolor{yellow}M\\ \cline{2-10} 
&\multirow{2}{*}{Hijacking}    & Vocabulary               &7.5&5&7.5&7.5&5&6.5& \cellcolor{yellow}M\\ \cline{3-10} 
&& Square                  &5&10&5&7.5&10&7.5&\cellcolor{orange}H\\ \hline
\multirow{8}{*}{\rotatebox[origin=c]{90}{\begin{tabular}[c]{@{}l@{}}\textbf{ Infrastructure-Based Attacks} \end{tabular}}} &\multirow{2}{*}{\begin{tabular}[c]{@{}l@{}}Unbounded \\ Consumption\end{tabular} }     & Denial of service            &10&7.5&7.5&10&7.5&8.5& \cellcolor{orange}H  \\ \cline{3-10} 
& & Denial of wallet           &10&7.5&7.5&10&7.5& 8.5& \cellcolor{orange}H\\ \cline{2-10} 
&\multirow{2}{*}{\begin{tabular}[c]{@{}l@{}}Inadequate \\ Sandboxing\end{tabular} }      & Environmental segregation              &7.5&5&5&7.5&7.5&6.5& \cellcolor{yellow}M \\ \cline{3-10} 
& & System-level interaction          &7.5&5&5&10&5&6.5&\cellcolor{yellow}M \\ \cline{2-10} 
& Access Control& &10&5&7.5&7.5&5&7&\cellcolor{orange}H \\ \cline{2-10}
& Supply chain & &10&5&5&10&5&7&\cellcolor{orange}H \\ \cline{2-10}
&\multirow{2}{*}{Side-Channel}      & Timing attacks              &5&7.5&10&10&10&8.5&\cellcolor{orange}H  \\ \cline{3-10} 
& & Power analysis          &7.5&7.5&7.5&10&7.5&8& \cellcolor{orange}H \\ \cline{2-10} 
& \begin{tabular}[c]{@{}l@{}}Server-Side \\Request Forgery \end{tabular} & &7.5&5&7.5&10&7.5&7.5& \cellcolor{orange}H \\ \cline{2-10}
 & \begin{tabular}[c]{@{}l@{}}Remote Code \\ Execution\end{tabular} & &10&5&7.5&10&7.5&8& \cellcolor{orange}H \\ \cline{2-10}            & Ransomware  & &10&5&5&10&5&7& \cellcolor{orange}H \\ \hline                 
\end{tabular}
\end{table*}

\subsection{DREAD-based risk assessment of LLMs Attacks}

\begin{figure}[!h]
   \centering
    \caption{Risk severity levels adopted from NIST}
\includegraphics[width=0.5\textwidth]{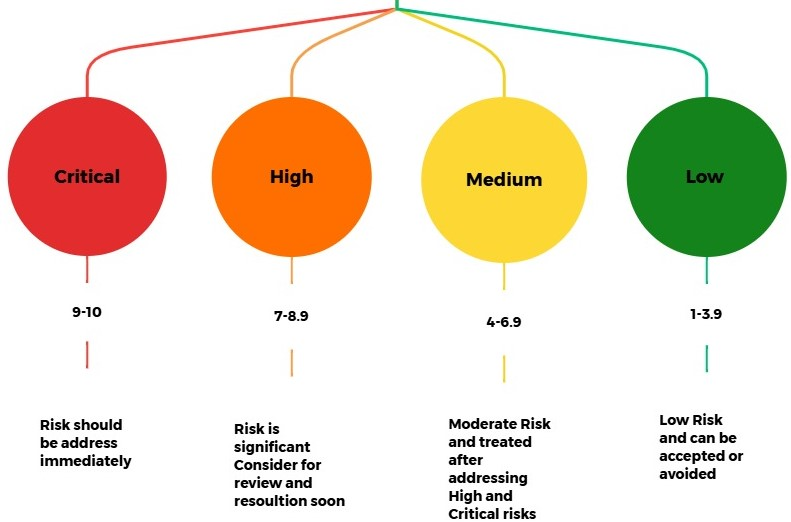}
  \label{fig:risk} 
%\vspace{-15pt}
     \end{figure}

We now show an application of the  generalised LLM attack taxonomy with the DREAD model for assessing LLM-based risk in the educational sector. In order to do so, each DREAD category has been evaluated and scored for each attack. It must be noted that these scores are provided for generally known and accepted use cases of \ac{LLMs} in various organisational aspects of an educational institution. It is likely that in a specific context and a specific organisation, the scores will vary. 
Table~\ref{educationtable} shows the overall risk score and risk severity level for each type of the attack categorised in our proposed taxonomy (Section.~\ref{sec:tax}) in the context of education sector. The overall risk scores are calculated using the average of the DREAD categories' scores (discussed in Section ~\ref{dreadmodel}). The risk severity levels (Figure~\ref{fig:risk}) have been adopted from the levels specified by NIST \cite{nvdd}. 

% \noindent\textbf{Low (1–3.9)}: Low risk to educational infrastructure and data and risk could be accepted or avoid.

% \noindent\textbf{Medium (4-6.9)}: Moderate risk to educational infrastructure and data; risk could be treated after addressing high and critical risks.

% \noindent\textbf{High (7-8.9)}: Risk is significant and should consider for review and resolution soon.

% \noindent\textbf{Critical (9-10)}: Risk should be address immediately.
Based on the security risk levels, it has been observed from the  Table~\ref{educationtable} that four attacks are the most critical risks for the education sector. Thirty-two (32) attacks out of 50 attacks targeting \ac{LLMs} are considered as the high risks within the education sector. None of the attack is at low level, and the remaining attacks are the medium risks. In the rest of this section, we discuss some examples to understand the assigned score.

\subsubsection{Ransomware Attack: A  high risk infrastructure attack in the education sector}

The calculated DREAD risk score and level show that the ransomware attack  (Section.~\ref{sec:taxinfr}) has high risk in the education sector (Table~\ref{educationtable}). The following description illustrates that a ransomware attack results in high damage and impacts a large number of stakeholders. However, as the attack is technically sophisticated compared to others, it is challenging to reproduce, exploit, and discover. Overall, this results in a high risk rating. If an attacker employs ransomware-as-a-service, the technical skills required are significantly lower, making the attack more easily reproducible. This aspect increases the severity of the ransomware threat. For instance,  if we assigned reproducibility criteria a value of 10 due to the ease, the overall average risk score will be 9.5,  making ransomware a critical threat for an organisation to focus on. However, in the following paragraphs, we have considered the scenario where attackers are executing the attack independently.  

\noindent\textbf{Educational Institution Reputational Damage}: A successful ransomware attack on \ac{eLLMs} could significantly damage the educational institution’s reputation among students, employees, parents, and wider internal and external stakeholders, resulting in the loss of public trust. This attack negatively impacts enrollments, academics, funding opportunities, partnerships, exchange programs, and campus operations. 

\noindent\textbf{Educational Operational Disruption}: A successful attack on LLMs could disrupt learning activities or academic operations. For example, access to the learner-facing LLMs- educational interactive resources like virtual tutors or agents, learning management systems (Moodle, Blackboard), virtual classrooms, and library chatbots become restricted due to the encryption, resulting in the cancellation of classes, exams, delayed feedback or results, and access to research materials and directly impact students’ futures. 

Similarly, the unavailability of educator-facing LLMs, which could support the teachers in providing writing analytics, smart content or assessment generation, personalized assessments, and automated grading and evaluation of exams, results in academic interruptions. 

Regarding unresponsiveness of institutional support \ac{LLMs} disrupt the fundamental academic operations such as student engagement, academic integrity, improvements in student retentions, teachers' evaluations, diagnosing strengths or gaps in student knowledge. Ransomware could target \ac{LLMs} used in research by corrupting the datasets or publicly available information.

\noindent\textbf{Data Breach of Affected Users}: Many \ac{LLMs} integrated into education infrastructure such as databases, servers hold critical and sensitive data, such as grades, personal information, financial information or even behavioral patterns. Ransomware attacks compromise the confidentiality and integrity of the data being encrypted. Similarly, these attacks result in intellectual property loss, either through theft or encryption,  by targeting research papers, lesson plans, and private educational content.

\noindent\textbf{Financial Losses}: Educational institutions hit by ransomware may be forced to pay ransom to decrypt the data or resume access to the \ac{LLMs}. Even if the ransom is not paid, institutions must incur the recovery cost for data restoration or system reconfiguration. Besides that, institutions have to pay penalties in case of data breaches.

\subsubsection{Content Manipulation Attack: A  high risk in the education sector } A content manipulation attack is the sub-type of manipulation attack discussed in Section.~\ref{sec:taxmodel}. In an educational environment, accurate and unbiased information is significant to maintain the trust of students, researchers and the academic staff. Therefore, this attack possesses a high risk in education.  

\noindent\textbf{Misinformation- Misleading, Incorrect or Biased Output and Disinformation}: An attacker (disgruntled student or staff member, script kiddie (novice hacker to gain recognition, and hacktivist) could manipulate educational content like self-learning materials, course content, course planners, research papers, e-books, or even exam material, spreading incorrect or fake information to students and deteriorating the quality of education. For example, institutions use \ac{eLLMs} as an intelligent tutoring system or virtual tutors to guide the students' online learning and to answer students' questions \cite{taneja2024jill}; manipulated content could mislead students, causing confusion and substandard student performance. Similarly, an attacker could craft eLLM's input with biased data to produce biased outputs, so there could be a discrepancy in student's knowledge with the market requirements. Also, \ac{eLLMs} used for generating research findings or insights could result in false or biased research outputs or poor research quality due to content manipulation impacting scientific or academic progress. 

Also, within the context of administration, attackers could alter the LLM's use for examination purposes. For instance, attackers could change exam paper content or manipulate grading criteria, or exam planners, resulting in unfair examinations and loss of academic integrity. Similarly, \ac{eLLMs} involved in student recruitment could impact the financial sustainability if the tuition fee pricing being varied by the attacker.   

In addition, the educator relies on third-party LLM, which an attacker compromises to produce biased output, to create content according to the approved course outline. The biased output in the teaching content could mislead the students and indirectly impact teacher's creditability.

\noindent\textbf{Exploitation/Risk of Overreliance on Technology:} Educators, learners, and administrations depend on \ac{LLMs} systems for educational content generation and decision-making without checking their facts (checking process in educational environments).  Attackers can exploit this ignorance and launch content manipulation attacks to integrate incorrect data into the model, which could impact critical decisions or processes,  result into internal risks to ensure the high-level quality educational services, and public embarrassment for the institution.

\subsubsection{Token Smuggling Attack: A  critical risk in the education sector}  Token smuggling attacks can be launched on \ac{eLLMs} for the creation of harmful or inappropriate learning materials or contents, gaining an  access to block or restricted contents, evading the cheating and plagiarism. Table~\ref{educationtable} shows that token smuggling attack has critical severity. The following are the reasons to emphasise that token smuggling is a critical risk for an educational institution, which should be mitigated proactively. 

\noindent\textbf{Harmful or Inappropriate Content Creation:} Attackers could use carefully selected words and phrases to generate harmful content on cyberbullying, trolling or even creating unsafe chemical experiments or violent scenarios on prohibited topics like improvising weapons, hacking networks, or accessing the dark web.

\noindent\textbf{Evasion of Plagiarism or Cheating:} Various \ac{LLMs} tools have emerged rapidly and have been utilized in educational institutions. However, token smuggling attacks enable the students to craft the responses by the instructions encoded in a manner that increases the risk that the educator may be unable to distinguish whether a student's writing is their work, resulting in unfair assessments. 

\noindent\textbf{Access to Block or Restricted Content:} Token smuggling can bypass these filters that \ac{eLLMs} have configured to block or restrict content. For example, students used smuggled tokens to trick \ac{eLLMs} into accessing exam questions or final grades or attendance records. Similarly, attacker could violate the intellectual property law and able to unblock the requests to websites to generate the content from copyright materials or retrieve the sensitive or confidential educational information. 

\subsection{Safeguards for Risk Mitigation in Education Sector}

The previous sections identified and characterised \ac{LLMs}-based attacks (on model directly or on infrastructure) along with the attack vectors involved and their impacts in general. To answer RQ2, we provide a risk assessment criteria  to evaluate the severity of identified  attacks in the education sector using DREAD model. In the literature, various technical risk mitigation strategies have been suggested \cite{LLMsecurity2025,nganyewou2022threat,cui2024risk}. 

Within an educational environment, establishing the safeguards to mitigate and address the risks raised due to \ac{LLMs}-based attacks is paramount to ensure that \ac{LLMs} continues to maintain the integrity of educational experiences. Following are some strategies  that should be adopted by educational institutions for risk mitigation.

\subsubsection{Enforcement of \ac{eLLMs}-Usage Policy} There is a lack of comprehensive policies and guidelines about AI usage, including \ac{LLMs} in education \cite{su151612451}. Therefore, educational institutions should establish clear guidelines for using \ac{eLLMs} ethically and sensibly and enforce accountability. Auditing and monitoring are significant for analysing log interactions and detecting unusual insights such as an indication of smuggling attacks, content manipulations while prompting. The sensitive queries should require authentication and a strict code of conduct for the user. Also, the users should report harmful, unsensible content generation by \ac{eLLMs}. The policies should be transparent and fair, consider all the concerned educational departments, and involve collaboration between educators, policymakers, and student representatives. 

\ 

\subsubsection{Threat Modeling and Risk Assessment} Due to the complexity and novelty of \ac{LLMs}, threat modeling is an ongoing and structured approach to identify, assess, and prioritise the threats to them \cite{burabari2024threat}. In the context of the educational system, to minimise the harm to the learners, educators, and institutions themselves, \ac{eLLMs} models and infrastructure should be assessed regularly to identify the attack surfaces, vectors, impacts, attacker's motives, and model performances. Similarly, risk assessment could help educational institutions to quantify and prioritise the risks (based on their  potential impact and likelihood) associated with \ac{eLLMs} , enabling them to determine the appropriate risk strategies (mitigation, transfer, avoidance, or acceptance). Threat modeling and risk assessment are essential for educational institutions to allocate their resources and efforts effectively and efficiently, to create a strong institutional culture and awareness, to reduce uncertainty for the students and educators, and to help prevent future incidents. 

 \ 
 
\subsubsection{Rapid Training and Awareness} Rapid training and awareness is the most important strategy where the educator and administrative staff should get support and awareness regarding the misuse of \ac{eLLMs}. Due to digital education and students' diverse learning needs and interests, a one-size-fits-all approach will not be sufficient. Therefore, depending on the personalised reliance on the \ac{eLLMs}, effective integration of these tools required educators and administrative staff to understand and identify unusual prompts, misinform and disinform scenarios or contents, and set strict usage policies to minimise cheating and plagiarism issues. Students should also be made aware about the use of \ac{eLLMs} responsibly and the potential risks associated with \ac{LLMs}-based attacks. 

\

\subsubsection{Regular Security Updates, Patching and Response Plans} The educational \ac{LLMs} models and their underlying infrastructure should be continuously updated  to address newly discovered vulnerabilities by software providing runtime guardrails. The cyber-security analyst personnel in information technology services (ITS) department should continuously monitor and detect misuse of \ac{eLLMs} and their performances in educational environment and develop an incident response plan to follow if some damage has been occurred due to attacks on these \ac{eLLMs}.

\

\subsubsection{Implement Strong Access Controls}  \ac{eLLMs} should be secured using multi-factor authentication (MFA) and role-based access control (RBAC). Also, educational institutions relying on \ac{LLMs} should implement secure key management protocols, least privilege principles and defense-in-depth measures.  

\section{Conclusion and Future Directions}
\label{sec:conclusion}
\ac{LLMs} have emerged as important tools in various industries, including education, and are used to perform various language-based tasks. However, these models are susceptible to security attacks that can impact the model, the infrastructure, and the organisation. This paper investigated and introduced a general taxonomy to categorise sophisticated attacks to \ac{LLMs} based on their attack complexity, which will be useful for academic and industrial practitioners to secure the \ac{LLMs} against malicious actors. Notably, our proposed generalised taxonomy could also be applied to  other sectors such as health-care, finance and industrial automation. To show its applicability, we have applied the proposed taxonomy in education. We also assess the severity of these attacks in the education sector using the DREAD risk assessment model and suggest a few risk mitigation strategies to prevent the identified attacks. 

In the future, it will be interesting to simulate attack scenarios on existing \ac{LLMs}-integrated educational tools like Moodle to identify vulnerabilities, assess the severity, and suggest effective controls accordingly, such as enforcing an educational LLM usage policy. We could also incorporate other risk assessment frameworks in future research to enhance the comprehensiveness and robustness of the investigation. A detailed mitigation framework that maps defensive strategies to each identified risk and attack type will be proposed, providing actionable guidance for security practitioners and system developers. 

%Also, the security of \ac{eLLMs} is in its infancy, so, in the future, the attacks on \ac{eLLMs} and observable information will be expressed across educational institutions using Structured Threat Information Expression (STIX) framework to enable effective defense against evolving attacks in critical sectors. LLMs will be fine-tuned using explainable AI (XAI) to understand \ac{eLLMs} decision-making processes. Also, we will propose a framework for preventing ransomware and content manipulation attacks on \ac{eLLMs}.   
\subsection{\textbf{ Declarations}} \label{sec:declare}
  
\medskip
\noindent
\textbf{Ethical Approval}
 This article does not report any research with human participants or animals.
 
 \medskip
\noindent
\textbf{Informed consent}
 This article does not report any studies with human participants.

\medskip
\noindent
\textbf{Financial interests}
 The authors do not have any financial interests in this research.

\medskip
\noindent
\textbf{Conflict of Interest}
All authors declare that there is no conflict of interest.

 \bibliographystyle{unsrt} 
% % \item \emph{Article number in~reference examples:}\\
% % See \cite{b32,b33}.

% % \item \emph{Example when using et al.:}\\
% % See \cite{b34}.

% % \end{itemize}
\bibliography{LLMmain} 
% % \bibitem{RefJ}
% % % Format for Journal Reference
% % Author, Article title, Journal, Volume, page numbers (year)
% % % Format for books
% % \bibitem{RefB}
% % Author, Book title, page numbers. Publisher, place (year)
% % % etc
% \end{thebibliography}

\end{document}